\documentclass[twocolumn,prc,floatfix,showpacs,preprintnumbers,nofootinbib,%
superscriptaddress]{revtex4}
\usepackage{mathrsfs}
\usepackage{amssymb}
\usepackage{amsmath}
\usepackage{graphicx}
\usepackage{hyperref}

\newcommand{\beq}{\begin{eqnarray}}
\newcommand{\eeq}{\end{eqnarray}}
\newcommand{\be}{\begin{equation}}
\newcommand{\ee}{\end{equation}}

\newcommand{\nn}{\nonumber\\ }
\def\del{\partial}

\newcommand{\A}{A_{_A}}
\newcommand{\B}{A_{_B}}

\newcommand{\subA}{{_A}}
\newcommand{\subB}{{_B}}

\newcommand{\Acoul}{\mathcal{A}}

\newcommand{\ot}{\mathbf{0}_T}
\newcommand{\rt}{\mathbf{r}_T}

\newcommand{\xt}{\mathbf{x}_T}
\newcommand{\yt}{\mathbf{y}_T}
\newcommand{\itt}{\mathbf{i}_T}
\newcommand{\jt}{\mathbf{j}_T}
\newcommand{\pt}{{\mathbf{p}_T}}
\newcommand{\ktt}{k_T} 
\newcommand{\qt}{\mathbf{q}_T}
\newcommand{\kt}{\mathbf{k}_T}

\newcommand{\nabt}{\boldsymbol{\nabla}_T}
\newcommand{\At}{\mathbf{A}_T}

\newcommand{\ud}{\, \mathrm{d}}

\newcommand{\tr}{\, \mathrm{Tr} \, }

\newcommand{\nc}{{N_\mathrm{c}}}

\newcommand{\half}{\frac{1}{2}}
\newcommand{\hc}{\mathrm{\ h.c.\ }}

\newcommand{\nr}[1]{(\ref{#1})}

\newcommand{\ra}{R_A}

\newcommand{\gev}{\ \textrm{GeV}}

\newcommand{\qs}{Q_\mathrm{s}}
\newcommand{\qsa}{Q_\mathrm{sA}}
\newcommand{\qsp}{Q_{\mathrm{s}p}}

\newcommand{\as}{\alpha_{\mathrm{s}}}

\newcommand{\cN}{\mathcal{N}}
\newcommand{\cNt}{\widetilde{\mathcal{N}}}

\newcommand{\raa}{R_{AA}}
\newcommand{\rpa}{R_{pA}}
\newcommand{\rab}{R_{AB}}

\newcommand{\fig}{Fig.~}
\newcommand{\figs}{Figs.~}
\newcommand{\eq}{Eq.~}
\newcommand{\se}{Sec.~}
\newcommand{\eqs}{Eqs.~}

\begin{document}

\author{J.-P. Blaizot}
\affiliation{
Institut de Physique Th\'eorique (URA 2306 du CNRS), B\^at. 774, CEA/DSM/Saclay, 91191 Gif-sur-Yvette, France
}
\author{T. Lappi}
\affiliation{
Department of Physics, %
 P.O. Box 35, 40014 University of Jyv\"askyl\"a, Finland
}
\affiliation{
Helsinki Institute of Physics, P.O. Box 64, 00014 University of Helsinki,
Finland
}
\author{Y. Mehtar-Tani}
\affiliation{
Departamento de F\'isica de Part\'iculas and IGFAE,
Universidade de Santiago de Compostela, E-15706 Santiago de Compostela,
Galicia, Spain
}

\title{
On the gluon spectrum in the glasma
}


\begin{abstract}
We study the gluon distribution in nucleus-nucleus collisions in the
framework of the Color-Glass-Condensate. Approximate analytical solutions
are compared to  numerical solutions of the non-linear Yang-Mills equations.
We find that the full numerical solution can be well approximated 
by taking the full initial condition of the fields in Coulomb gauge
and using a linearized solution for the time evolution. We also compare
$\ktt$-factorized approximations to the full solution.
\end{abstract}

\maketitle

\section{Introduction}

In the Color Glass Condensate (CGC) framework (for reviews
see~\cite{Iancu:2003xm,Weigert:2005us}), particle
multiplicities are dominated by classical gluon dynamics.  The spectra
of gluons produced in the collision of two heavy nuclei at high energy
is determined~\cite{McLerran:1994ni,McLerran:1994ka,McLerran:1994vd}
completely by the classical glasma~\cite{Lappi:2006fp} gauge field
$A^\mu$ that is the solution of the classical Yang-Mills (CYM)
equations of motion:
\be
[D_\mu, F^{\mu\nu}]=J^\nu,
\ee
where $D_\mu=\del_\mu-igA_\mu$ is the covariant derivative and
$F_{\mu\nu}=\del_\mu A_\nu -\del_\nu A_\mu -ig[A_\mu,A_\nu]$. The
current $J^\mu$, which must be covariantly conserved ($[D_\mu,J^\mu]=0$),  
describes the colliding nuclei. At lowest order in the color
 densities of the nuclei it reads
\be
J^+=\rho_\subA(x^+,\xt),\; J^-=\rho_\subB(x^-,\xt),\; J^i=0,
\ee
with $x^{\pm}=(t\pm z)/\sqrt{2}$. This current reflects the kinematics
 of two nuclei $A$ and $B$ moving in
the $-z$ and $+z$ direction, respectively, at almost the speed of light. 
With a specific
gauge choice, for instance with the light-cone gauge or the Fock-Schwinger
gauge, the conserved current can be fully determined, allowing us to
solve the CYM equations on the light-cone.  The
problem of gluon production reduces then to an initial value problem
of solving the CYM equations  $[D_\mu,F^{\mu\nu}]=0$
in the forward light-cone, $x^\pm \geq 0$. The  initial condition for
the gauge field can be determined explicitly by solving the 
CYM near the light-cone, i.e., at proper time $\tau=\sqrt{2x^+x^-}=0^+$. 

An often used approximation for computing gluon production in heavy
ion collisions is to express the spectrum as a convolution of the 
unintegrated gluon distribution functions, leading to the so-called 
$\ktt$-factorization, which holds exactly in the dilute  ``pp'' 
(proton-proton) and the dilute-dense ``pA'' (proton-nucleus)
limits, that is,  when one or both of the color sources are 
 weak~\cite{Kharzeev:2003wz,Blaizot:2004wu,Gelis:2005pt,Gelis:2006tb}. 
In these cases the CYM equations can be linearized and solved 
explicitly\footnote{For a conjectured analytical solution in 
the fully nonlinear case see Ref.~\cite{Kovchegov:2000hz}.}.

In a recent work \cite{Blaizot:2008yb}
the spectrum of gluons has been obtained by 
solving the Yang-Mills equations near the light-cone and treating the 
evolution in time linearly. It has been  shown that 
this procedure yields the correct ``pA'' spectrum after
expanding the solution at first order in the weak 
proton sources,  provided one  chooses a gauge that removes the transverse
 pure gauge component of the fields of the nuclei when they are not interacting.
In the present paper, we investigate the accuracy of this approach
in the nucleus-nucleus case.
More precisely, the following questions will be addressed:
\begin{itemize}
\item What is the role of the nonlinearities in the initial condition 
of the CYM calculations and in the time evolution for $\tau>0$? How
well can one approximate the full solution of the nonlinear equations
by one where the initial condition is treated exactly but the
time evolution is assumed linear?
\item How good an approximation is a $\ktt$-factorized expression for 
the gluon spectrum in the glasma? 
\end{itemize}
We will argue in the following that  a specific gauge,
namely the Fock-Schwinger gauge $A_\tau=0$ 
combined with a two dimensional transverse Coulomb gauge condition
$\del_i A_i = 0$ at $\tau=0^+$, allows us to minimize the magnitude of
the initial 
gauge fields and therefore to reduce the ``final state interactions,'' i.e., 
the non-linear dynamics inside the forward-light-cone. 
 We shall then, in \se\ref{sec:ktfact},
 discuss the range of validity of $\ktt$-factorization.

In order to answer the questions above, we shall compare approximate
solutions with the full numerical solutions of the CYM equations, obtained
using the method developed
in Refs.~\cite{Krasnitz:1998ns,Krasnitz:2000gz,Krasnitz:2001qu,Lappi:2003bi,Krasnitz:2003jw}.
As we shall discuss in \se\ref{sec:linearized}, the initial conditions
for the solution are given in terms of Wilson lines that, in the CGC,
are random SU(3) matrices, correlated in the transverse plane
over distances of order $1/\qs$. 
 For our numerical studies we shall
generate these Wilson lines from an implementation of the McLerran-Venugopalan
 (MV) model~\cite{McLerran:1994ni,McLerran:1994ka,McLerran:1994vd} 
(for more detail see Refs.~\cite{Lappi:2003bi,Lappi:2007ku,Lappi:2009xa}),
where the color densities are treated as random variables with 
a Gaussian distribution of variance 
\begin{multline}\label{eq:variance}
\langle \rho^a_\subA(x^+,\xt) \rho^b_\subA(y^+,\yt)\rangle
=
\\
g^2\mu^2(x^+) \delta^{ab}\delta(x^+-y^+)\delta(\xt-\yt),
\end{multline}
for nucleus A and similarly for nucleus B. 
Here, $g^2 \mu^2(x^+)$ stands for the three-dimensional (squared)
color charge density.
After the averaging over the color densities, the observable will be only function 
of the two dimensional density of (squared) color charge
\be
g^2 \mu_\subA^2 = \int_{-\infty}^{+\infty} \ud x^+ g^2 \mu^2(x^+).
\ee
We denote the value integrated over the longitudinal coordinate $x^\pm$
by $\mu^2$ without an explicit dependence on  $x^\pm$.
The saturation scale $\qs$ is is proportional to 
(and numerically of the same order of magnitude as) 
$g^2\mu_\subA$. The relevant values are of order 
$g^2\mu_\subA = 1 - 2\gev$ for RHIC and $2 - 3\gev$ for the LHC. 
Note that in the convention used here $\mu^2_\subA$
is a number density of charge carriers and $g^2\mu_\subA^2$ the
squared charge density.
In the saturation regime $g^2\mu_\subA^2 \sim 1/\as$ 
so that the saturation scale $\qs \sim g^2\mu_\subA$
is parametrically independent of the coupling.
For a discussion 
of the relation between the MV model parameter $g^2\mu$ and the 
saturation scale $\qs$ as measured in DIS experiments we refer the reader
to Refs.~\cite{Lappi:2007ku,Lappi:2009xa}.
We emphasize that the main purpose of this paper is not to
compare directly to experimental data, but to compare different
approximations for the gluon spectrum in the glasma corresponding to the
same distribution of color charges (or more properly Wilson lines,
as we shall see shortly). Our main results would therefore be independent
of the details of the color charge density distribution,  and equally 
valid for one that would match more closely, e.g., the solution of
the JIMWLK/BK evolution equations.
Note however, that
although we do not attempt to choose values for the parameters of our
calculation to give an accurate best estimate for RHIC or the LHC, 
we shall stay in the phenomenologically relevant range.

\section{Linearized approximation of Yang-Mills equations}
\label{sec:linearized}
In this section we describe our linearized approximation 
for solving the CYM equations.  

We work in the the Fock-Schwinger (FS) gauge $\tau A_\tau \equiv  x^+ A^-+x^- A^+=0$.
 In Ref.~\cite{Blaizot:2008yb} the discussion was formulated in terms of the light
cone (LC) gauge. Appendix C of Ref.~\cite{Blaizot:2008yb} shows how both gauge 
conditions lead to the same functional form of the spectrum in the linearized approximation,
\eq\nr{eq:spectrum2} of the present paper.
The remaining longitudinal component of the gauge field 
is most naturally parametrized by the component  $A^\eta$ orthogonal to $A_\tau$,
namely as $A^{\pm} = \pm x^{\pm}A^\eta$, where $\eta=\frac{1}{2}\ln (x^+/x^-)$ 
is the spacetime rapidity. 

The initial condition for the field on the light-cone is
given by~\cite{Kovner:1995ts,Kovner:1995ja}
\begin{eqnarray}\label{eq:initcond}
\left. A^i \right|_{\tau=0^+} &=& \A^i + \B^i, \nn
\left. A^\eta \right|_{\tau=0^+} &=& \frac{ig}{2}[\A^i,\B^i],
\end{eqnarray}
where the LC gauge fields for the individual nuclei are given by
\begin{equation}\label{eq:infields}
 \A^i= -\frac{1}{ig} U^\dag\partial^i U, \; \B^i= -\frac{1}{ig}
 V^\dag\partial^i V.
\end{equation}
The fundamental degrees of freedom characterizing the
CGC wavefunctions of the individual nuclei are the
\emph{Wilson lines} $U(\xt)$ and $V(\xt)$,
\be \label{eq:UWline}
U(\xt)\equiv {\cal P}_+ \exp\left[ig\int_{-\infty}^{+\infty} \ud z^+
\frac{1}{\nabt^2}\rho_\subA(z^+,\xt)\right],
\ee
and 
\be\label{eq:UWline2}
V(\xt)\equiv {\cal P}_- \exp\left[ig\int_{-\infty}^{+\infty} \ud z^-
\frac{1}{\nabt^2}\rho_\subB(z^-,\xt)\right],
\ee
where $\rho_\subA(x^+,\xt)$ and $\rho_\subB(x^-,\xt)$  are the color charge densities 
of the nuclei. 
In the numerical implementation of the MV model these Wilson lines are 
constructed as
\begin{equation}\label{eq:uprod}
U(\xt)  = \prod_{k=1}^{N_y} \exp\left\{ -i g \frac{1}{\nabt^2}
\rho_k^{\subA,\subB}(\xt) \right\},
\end{equation}
where the color charges are Gaussian variables with the variance
\begin{equation}\label{eq:discrsrc}
\left\langle \rho^a_k(\xt) \rho^b_l(\yt) \right\rangle =
 \delta^{ab}  \delta^{kl}  \delta^2(\xt-\yt)
\frac{g^2 \mu^2}{N_y}.
\end{equation}
The indices $k,l=1\dots N_y$ represent a discretization of the longitudinal 
direction into $N_y$ small steps; the continuum limit corresponding to  
\eqs\nr{eq:UWline} and~\nr{eq:UWline2} is achieved for  $N_y\to \infty$ at 
constant $g^2\mu$.
Some kind of infrared regulator is needed in order to invert the Laplacian operator
$\nabt^2$. One possibility is to replace $\nabt^2$ in \eq\nr{eq:uprod}
by $\nabt^2 + m^2$, with a parameter $m$ chosen so that $m \ll \qs$. Another
possibility is to set the $\kt=\ot$-mode of $\rho$ to zero, i.e. to
impose total color neutrality on the source. This corresponds
to an infrared cutoff given by the size of the system; this is the procedure
we use in what follows in the cases where we take $m=0$. 
As noted before, our purpose is to compare different methods to obtain the
gluon spectrum from the same distribution of Wilson lines. Therefore
the important comparison in this context is between different approximations
for the same values of $N_y$, $m$ and $g^2\mu$.  
We shall not carry out a 
a systematic study of the dependence on these parameters 
separately (see Ref.~\cite{Lappi:2009xa} for a more detailed analysis),
except for the specific case of the dependence on $N_y$ in \se\ref{sec:raa}.

Note that when the fields are boost invariant, the gauge condition
$A_\tau=0$ leads to $\del_+A^+ + \del_-A^-= 0$.  This gauge, as well as 
the LC-gauge,  does not fix completely the gauge field,
and we may exploit the remaining gauge freedom in order to 
simplify the calculation.
Imposing the restriction that we want to stay within the
Fock-Schwinger gauge $A_\tau=0$ forbids $\tau$-dependent gauge
transformations. We also want to preserve the explicit boost
invariance of the field configurations and therefore we do not want to
perform $\eta$-dependent transformations. This leaves us the freedom of
gauge changes  that depend (in the region
$\tau>0$) only on the transverse coordinates.
We denote the transformed field by $\Acoul$:
\be\label{eq:gaugerotation}
\Acoul^\mu = \Omega A^\mu\, \Omega^\dag-\frac{1}{ig} \Omega\del^\mu
\Omega^\dag.
\ee
The initial conditions in the new gauge are
\beq
\left. \Acoul^\eta\right|_{\tau=0^+} &=&\frac{ ig}{2}\Omega [\A^i,\B^i]\Omega^\dag\nn
\left. \Acoul^i\right|_{\tau=0^+} &=& \Omega(\A^i+\B^i)\,\Omega^\dag-\frac{1}{ig} \Omega\del^i\Omega^\dag,
\label{eq:gaugerotation2}
\eeq
where $\A^i$ and $\B^i$ are given by \eq\nr{eq:infields}.
As we shall see, we can choose $\Omega$ so as to 
reduce final state interactions 
and treat the evolution of the fields
 after the collision in a linear approximation. As recalled earlier, a motivation 
for this strategy is the fact that, in a 
suitable gauge, treating the final state dynamics to lowest order in the
field gives the exact solution in the case where one of the projectile
is dilute and the other one dense (the ``pA'' case). 

We shall return to the determination of $\Omega$ later, and first 
review the linearized solution, following
Ref.~\cite{Blaizot:2008yb}. The linearized equations of motion, for $\tau>0$, are 
\beq
\square \Acoul^i &=&-\del^i \del^j \Acoul^j,  \label{eq:FSYMtr} \\
\square \Acoul^\pm&=&-\del^\pm\!\del^j \Acoul^j \label{eq:FSYMl}.
\eeq
These are to be solved with the initial conditions given in \eq\nr{eq:gaugerotation2}.
The second equation \nr{eq:FSYMl}, combined with the gauge condition and
boost invariance, leads to
\be \label{eq:coulcons}
\square \left[ \del^+  \Acoul^-+\del^- \Acoul^+\right]
=2\del^+\del^-\del^j \Acoul^j 
=0,
\ee
which states that
the divergence of the transverse field is conserved.
This allows us to rewrite the Yang-Mills equations in a form that
makes the boundary conditions explicit. Following the same steps as in
Ref.~\cite{Blaizot:2008yb}, we obtain 
\beq\label{eq:FSYM2}
\square \Acoul^i &=&2\delta(x^+)\delta(x^-)\left. \Acoul^i\right|_{\tau=0^+}
\\  \nonumber && 
 - \theta(x^+) \theta(x^-) \partial^i \del^j \left. \Acoul^j \right|_{\tau=0^+} , \\
\square \Acoul^+&=&\delta(x^-)\theta(x^+)\left. \Acoul^\eta \right|_{\tau=0^+},\\ 
\square \Acoul^-&=& - \delta(x^+)\theta(x^-) \left. \Acoul^\eta \right|_{\tau=0^+}.
\eeq
Fourier transforming these equations we get 
\beq\label{eq:FTsolution}
&&-k^2 \Acoul^i(k)=2\left(\delta^{ij}-\frac{k^ik^j}{2k^+ k^-}\right) \left. \Acoul^j(\kt) \right|_{ \tau=0^+}, \\
&&-k^2 \Acoul^+(k)=-\frac{i}{k^-} \left. \Acoul^\eta(\kt) \right|_{ \tau=0^+},\\
&&-k^2 \Acoul^-(k)= \frac{i}{k^+} \left. \Acoul^\eta(\kt) \right|_{ \tau=0^+}.
\eeq

The gluon spectrum is then 
computed with the help of the reduction formula:
\be\label{eq:spectrum1}
4\pi^3E\frac{\ud N}{\ud^3{\bf k}}=\sum_\lambda |{\cal M}_\lambda|^2,
\ee
with the production amplitude ${\cal M}_\lambda=
\lim\limits_{k^2\rightarrow 0}k^2\epsilon^{\lambda}_\mu \Acoul^\mu(k)$ 
for a gluon of polarization $\lambda$. 
By using the completeness
relation $\sum\limits_\lambda \epsilon_\mu^\lambda
(\epsilon_\nu^\lambda)^*= - g_{\mu\nu}$, one finally gets~\cite{Blaizot:2008yb}:
\begin{multline}\label{eq:spectrum2}
\frac{\ud N}{\ud y \ud^2 \kt}= \frac{1}{(2 \pi)^2}\frac{1}{\pi \kt^2} 
\Big\langle
 \left| \kt\times {\bf \cal A }(\kt) \right|^2 
\\
+\left|  \Acoul^\eta(\kt)\right|^2 \Big\rangle_{\tau=0^+},
\end{multline}
where the cross product stands for
$\kt\times {\bf \cal A} \equiv \epsilon^{ij} k^i \Acoul^j$.

Let us now return to the choice of the gauge transformation 
$\Omega$ in \eq\nr{eq:gaugerotation}. The choice introduced
in Ref.~\cite{Blaizot:2008yb} (see Appendix C of ~\cite{Blaizot:2008yb}
for the explicit
derivation) was to take either $\Omega = UV$ or $\Omega = VU$. 
It was shown in Ref.~\cite{Blaizot:2008yb} that  this choice 
reproduces correctly both the ``pA'' and ``Ap'' limits (because the
gauge is not symmetric in the two nuclei one has to study the
two cases separately). In general, any choice that would gauge 
away the pure transverse fields of the nuclei, $A^i_{\subA}$ and $A^i_{\subB}$, 
before the collision, leads to the correct answer for ``pA''. 
Thus we look for $\Omega$ that verify the following  
boundary conditions:  when $x^-<0$, i.e., in the absence of nucleus B,
$\Omega\equiv U$ and when $x^+<0$, i.e., in the absence of nucleus A,
$\Omega\equiv V$. 

With the explicit expression
$\Omega=VU$ one can directly perform the gauge rotations of the 
initial conditions in \eq\nr{eq:gaugerotation2} to get\footnote{
These are \eqs(2.73) and~(2.74) of Ref.~\cite{Blaizot:2008yb}
written in the fundamental representation.}
\begin{eqnarray}
\label{eq:lcinitc1}
\left. \Acoul^i \right|_{\tau=0^+} &=& V \left( U \B^i U^\dag - \B^i \right) V^\dag 
\\
\label{eq:lcinitc2}
\left. \Acoul^\eta \right|_{\tau=0^+} &=& V \left( \del^i U \B^i U^\dag +
U \B^i \del^i U^\dag \right) V^\dag.
\end{eqnarray}
The expression for the spectrum that results from inserting 
\eqs\nr{eq:lcinitc1} and~\nr{eq:lcinitc2} in \eq\nr{eq:spectrum2}
was obtained in Ref.~\cite{Blaizot:2008yb}.

We now look for a  gauge where the linearized
evolution for $\tau>0$ works best, that is, we 
look for a gauge that minimizes the value of the gauge potential
$\Acoul$. With our restriction of boost invariance and the Fock-Schwinger
condition $A_\tau=0$ a natural way to do this is to minimize the transverse 
components of $\Acoul$. They are the ones in which the large unphysical 
pure gauge contributions show up, as can be explicitly seen in the
computation of Refs. \cite{Kovner:1995ts,Dumitru:2001ux}.
A convenient prescription for this is to 
minimize the functional $\int \ud^2 \xt |\At (\xt,\tau=0^+)|^2$; this is 
achieved by the 2-dimensional Coulomb gauge $\del^i \Acoul^i = 0$.
The Coulomb gauge condition is conserved by the linearized equations
of motion (see \eq\nr{eq:coulcons}), so it is equivalent to impose
 it at $\tau=0^+$ or at 
a larger $\tau$. The full equations of motion, on the other hand,
do not conserve the gauge condition. Thus in the numerical computation
the gauge condition is imposed at the end of the evolution, when the 
gluon spectrum is calculated.

In order to impose the condition $\del^i A^i = 0$ we
must then find the gauge transformation 
$\Omega$ as a solution of the equation 
\be\label{eq:gaugelink}
\del^i\left(\Omega (\A^i+\B^i)\,\Omega^\dag-\frac{1}{ig} \Omega\del^i\Omega^\dag\right)=0.
\ee
When $x^-<0$,
$\Omega\equiv U$ and when $x^+<0$, $\Omega\equiv V$ as required to reproduce the ``pA'' spectrum. In this case the pure transverse initial fields get
rotated leading to pure longitudinal fields, i.e.,
\beq
&& \Acoul^+=\frac{1}{\nabt^2} \rho_\subB(x^-,\xt), 
\; \Acoul^-=\Acoul^i=0,\;\text{for} \; x^+<0,\nn 
&& 
\Acoul^-=\frac{1}{\nabt^2} \rho_\subA(x^+,\xt), 
\; \Acoul^+=\Acoul^i=0,\;\text{for} \; x^-<0.\nn
\eeq
In the strong field case we are not able to solve \eq\nr{eq:gaugelink}
for $\Omega$ analytically (or even to show formally that a unique solution exists).
Finding the required gauge transformation numerically is, however,
not excessively difficult and has been a standard part of the
numerical CYM computations of the gluon 
multiplicity~\cite{Krasnitz:2001qu,Krasnitz:2000gz,Lappi:2003bi}.
 Indeed evaluating the approximate
formula \eq\nr{eq:spectrum2} in the Coulomb gauge defined by 
\eq\nr{eq:gaugelink} is significantly less demanding than solving the 
full time-dependence of the Yang-Mills equations.

In the following sections we shall  compare the approximate analytic
solution \eq\nr{eq:spectrum2} in the  two gauges discussed above,
$\Omega=VU$, and the Coulomb gauge choice, to the full numerical solution of the 
Yang-Mills equations. In addition to giving physical insight into the
(admittedly gauge dependent) question of the importance
of initial (meaning $\tau=0$) and final ($\tau \sim 1/\qs$ in this context)
state interactions in the glasma this may also constitute a 
starting point for further analytical studies.

\section{Linear and nonlinear final state dynamics}
\label{sec:gauges}

\begin{figure}
\includegraphics[width=0.48\textwidth]{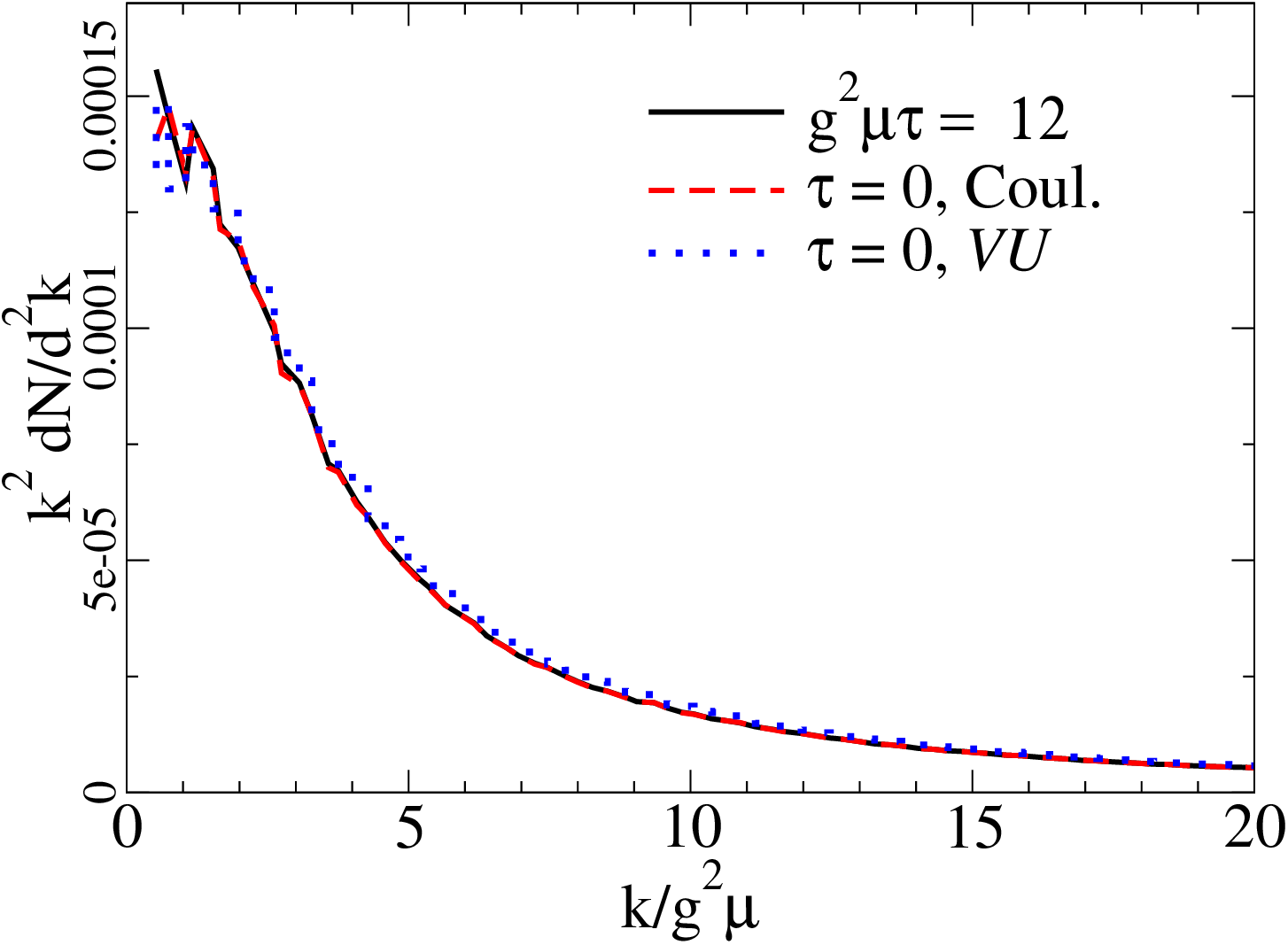}
\caption{$\Omega=VU$ gauge gauge and Coulomb gauge results compared 
to the full CYM result in the weak field regime with 
$g^2\mu = 0.2\gev$, $m=0.1\gev$, $N_y = 20$. 
The legend ``$\tau=0$'' refers to the spectrum being evaluated
using \eq\nr{eq:spectrum2} which expresses the multiplicity 
in terms of the fields at $\tau=0$. The full CYM result is
evaluated at $\tau = 12/g^2\mu$.
}
\label{fig:lccouldilute}
\end{figure}

\begin{figure}
\includegraphics[width=0.48\textwidth]{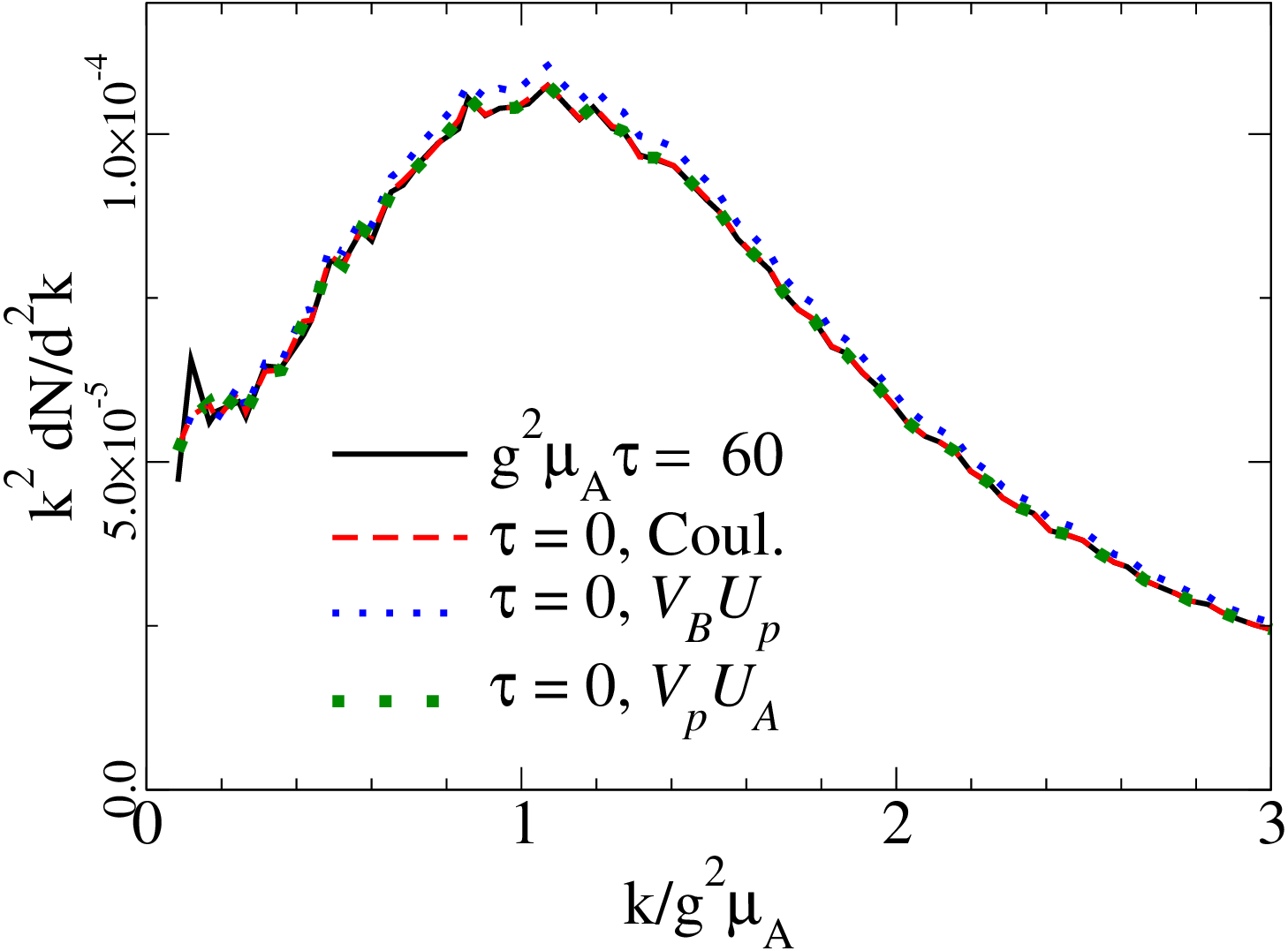}
\caption{$\Omega=VU$ gauge and Coulomb gauge results compared to the 
full CYM result in the ``pA'' case 
$g^2\mu_\subA = 1.25 \gev$,  $g^2\mu_p = 0.008 \gev$.
 There are two  $\Omega=VU$ gauge curves, corresponding to the 
cases where either the source
$A$ or  $B$ is taken to be weak, i.e. $V_\subB U_p$: source $A$ is the 
proton, and  $V_p U_\subA$: source $B$ is the proton.
It can be seen that in this limit of 
small $g^2\mu_p$ they are equivalent, but the $V_\subB U_p$  approximation 
approaches the dilute limit more slowly.
The CYM result is evaluated at  $\tau = 60/g^2\mu_\subA$. $N_y=20$ and $m=0$.}
\label{fig:lccoulpA}
\end{figure}

We start by comparing the results of the linearized evolution in the
 $\Omega=VU$ case and the 
Coulomb case in the dilute limit for both sources. 
Figure~\ref{fig:lccouldilute} shows the gluon spectrum
for $g^2\mu = 0.2\gev$, $N_y = 20$ and $m=0.1\gev$.
This situation is dilute enough to find a good agreement; 
the result mainly serves as a
check of the normalization in our numerical computation.
The uneven structure at small $\ktt$ in the full CYM calculation 
is an oscillation caused by the fact that the evolution is
stopped at a finite time $\tau$; see Appendix~\ref{sec:cymmulti}
for a more detailed discussion.

A slightly less trivial check is provided by the ``pA'' case. We write this in
quotation marks because  what we are really computing is the dilute--dense limit
where the smaller one of the two saturation scales is taken to be 
much smaller than the other, but also the weaker source is 
allowed to fill the whole transverse plane (of area $S_\perp = \pi\ra^2$) so that 
we need not worry about effects of the edge of the nucleus. Here also
several analytic calculations in the Coulomb~\cite{Dumitru:2001ux},
covariant~\cite{Kharzeev:2003wz,Blaizot:2004wu} and LC~\cite{Gelis:2005pt,Blaizot:2008yb} 
gauges have shown that the result can be written in a $\ktt$-factorized form, 
in which one needs to account for nonlinear interactions only in the initial condition.
A practical question is then of course how weak the ``proton'' source has
to be to be in this limit. 
Our numerical  
comparison is shown in \fig\ref{fig:lccoulpA}, where the Coulomb and $\Omega=VU$ gauge 
approximations are seen to agree with the full numerical computation. Note that
because the $\Omega=VU$ gauge prescription, \eqs\nr{eq:lcinitc1} and ~\nr{eq:lcinitc2},
is asymmetric in the nuclei A and B, one
has to look at the ``pA'' and ``Ap'' limits of the approximation separately.
In \fig\ref{fig:lccoulpA} these are denoted by $V_\subB U_p$ (source $A$ is the proton)
and  $V_p U_\subA$ (source $B$ is the proton).
Inspecting
the figure closely one can see that the $V_\subB U_p$ curve is slightly further away
from the full CYM result. This signals the fact that the limit of nucleus 
$B$ being dilute converges towards the linear regime faster than the limit
of $A$ being dilute, which we have verified for less asymmetric values
of $g^2\mu_\subA$ and $g^2\mu_p$. In other words, for the ``marginal pA'' case
in which the proton source is weak but not infinitesimal,  the gauge 
transformation $\Omega = V_p U_\subA$ gives a slightly better approximation 
to the full result than $\Omega = V_\subB U_p$, where the Wilson line of the proton is
denoted with subscript $p$~\footnote{ To trace the origin of this small effect,
look at the analytical expressions for the
``pA''  and ``Ap'' limits, \eqs(3.91) and~(3.96), in Ref.~\cite{Blaizot:2008yb}.
In \eq(3.91) $\pt$ is parametrically large, $\sim \qsa$ and $\kt$ small $\sim \qsp$.
The longitudinal component of the gauge field 
$\tilde{\alpha}^i$ is proportional to the large momentum
$p^i$. In \eq(3.96) $\kt$ is large and $\pt$ small, and the longitudinal 
 component of $\tilde{\alpha}^i$ is proportional to the small momentum $p^i$.
Since the ``unphysical'' longitudinal polarization is smaller in (3.96), it is 
a better approximation.}.

\begin{figure}
\includegraphics[width=0.48\textwidth]{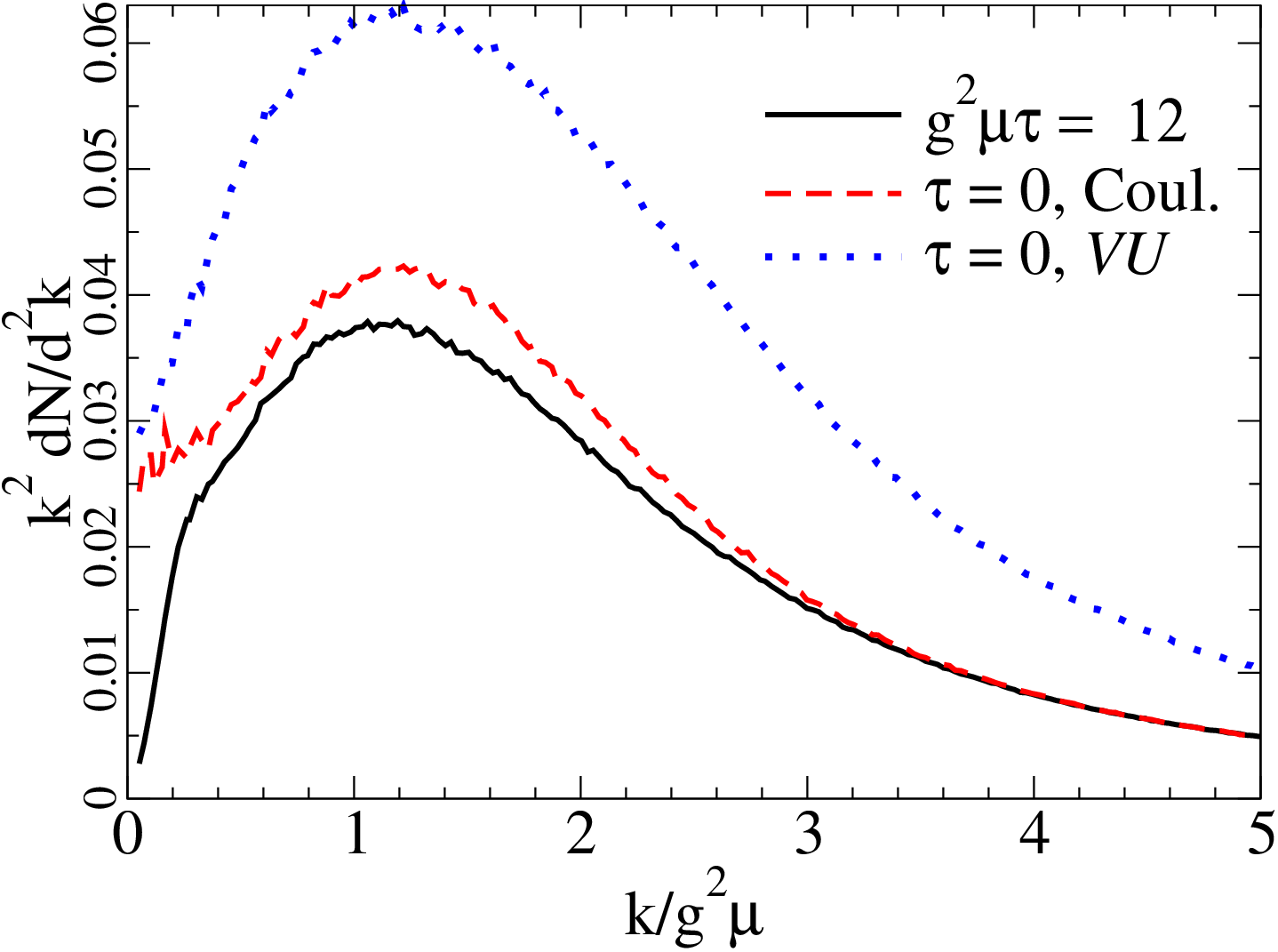}
\caption{Light-cone gauge and Coulomb gauge results compared to the full CYM result
in the saturated  strong field regime with $g^2\mu = 2\gev$, $m=0.1\gev$ $N_y=20$.
}
\label{fig:lccoul}
\end{figure}

Let us then turn to the same comparison for the nucleus-nucleus case when both 
sources are strong. The comparison between the linearized prescriptions
and the full result is shown in \fig\ref{fig:lccoul}. We can see that the 
linearized Coulomb gauge result is still relatively close to the full numerical
solution, but the one in $VU$-gauge starts to deviate from it significantly. 
After our discussion in \se\ref{sec:linearized} this behavior is easy to interpret.
The Coulomb gauge condition removes, practically by construction, 
the unphysical pure gauge component from the field. A linearized 
approximation works best in a 
gauge that minimizes the value of the gauge potential. The $VU$-gauge condition 
removes the pure gauge part of the field in the ``pA'' case, but not in the
``AA'' one, because the gauge transformation required to do this 
is more complicated. 
Let us comment in more details the comparison between the linearized 
Coulomb gauge result with the full CYM result. We distinguish typically 
three regions  in the spectrum: 
\begin{itemize}
\item The high momentum range, $\ktt/g^2 \mu> 3$. The two curves merge 
together, this is expected since both tend to the same limit $\sim 1/\ktt^2$
at high momentum.
\item The intermediate momentum region,  $0.2<\ktt/g^2 \mu<3$ , 
where the two curves follow the same trend and are peaked at
$\ktt/g^2 \mu \simeq 1.2$.
In this region, our analytical formula overestimates the 
full CYM result by   less than 15 \%.  
\item The low momentum region,  $\ktt/g\mu^2<0.2$. Here, we see that 
the two curves diverge. The full result tends to 0 whereas our
 approximate formula tends to a constant.
\end{itemize}
Note that the spectrum is multiplied by $\ktt^2$ in our figures, which focuses 
attention on the larger $\ktt$-part and minimizes the difference between the 
exact calculation and the approximate ones.
But the difference between the Coulomb 
gauge result and the full CYM result for the integrated multiplicity 
remains large although the spectra are close in shape. The difference
in the integrated multiplicity comes mostly from the infrared part of the spectrum.
Indeed  since the Coulomb-gauge spectrum is logarithmically
IR divergent it does not give a reliable estimate for
the integrated multiplicity. One may argue that, due
to the uncertainty in the CYM calculation
for $\ktt \lesssim 1/\tau$ the very small $\ktt$-part of the spectrum 
in \fig\ref{fig:lccoul} is is not very reliable either. Earlier numerical 
studies~\cite{Krasnitz:2000gz,Krasnitz:2001qu,Lappi:2003bi} have
shown, however, that the total gluon multiplicity is infrared finite. 
This IR-finiteness of the CYM result, presumably due to 
screening effects at small transverse 
momenta~\cite{Krasnitz:2000gz,Romatschke:2005pm,Romatschke:2006nk},
cannot be captured by a linearized treatment of the evolution for $\tau>0$.

\section{Nuclear modification factor}
\label{sec:raa}

\begin{figure}
\includegraphics[width=0.48\textwidth]{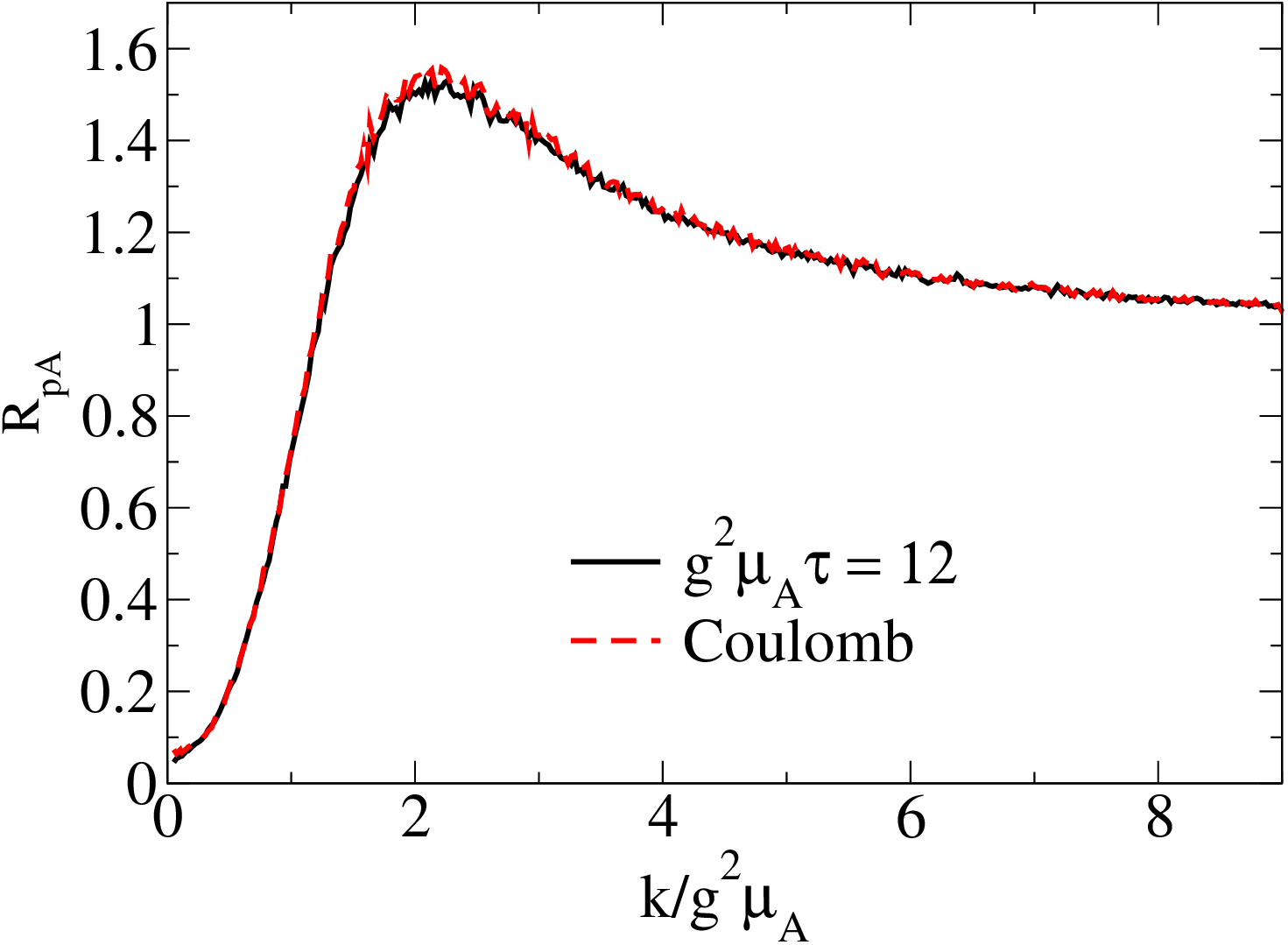}
\caption{
Nuclear modification factor of the gluon spectrum
in the saturated asymmetric ``pA'' case 
$\mu_\subA = 2\gev$, $g^2\mu_p = 0.32\gev$, $m=0.1\gev$ $N_y=500$.
}
\label{fig:RpA}
\end{figure}

We have shown in the previous section that an initial condition including 
all orders in the classical field in the Coulomb gauge followed by a linearized 
solution to the equations of motion gives a good approximation to
the full solution except at very small momenta.
 Let us now further illustrate  this point by computing
the nuclear modification factors of the gluon spectrum. These are defined
as the ratio of the spectrum of produced gluons to the corresponding
spectrum in the dilute ``pp'' case, normalized by the appropriate geometrical
 factor, the number of binary collisions $N_\textrm{coll}$, 
to get a quantity that should approach 1 at large transverse momenta.
In order to avoid complications with edge effects and
Coulomb tails of the gauge field extending outside
the nucleus~\cite{Krasnitz:2002ng,Krasnitz:2002mn,Lappi:2006xc} 
we compute the gluon spectrum in collisions
of two objects of the same size (filling the whole transverse lattice), but
with different saturation scales representing a proton or a nucleus.
We then compare the obtained gluon spectra per unit transverse area 
($S_\perp$) and 
normalize these using the expected large $\ktt$
behavior of the gluon spectrum as $\sim (g^2\mu_\subA)^2(g^2\mu_\subB)^2/\ktt^4$ to
get a quantity that approaches unity at large momenta.
Thus for a collision between two generic nuclei $A$ and $B$ our definition of
the nuclear  modification factor is
\be\label{eq:RAB}
\rab=\frac{\mu^4_p}{\mu^2_\subA\mu^2_\subB}
\left(
\frac{\ud N_{AB}}{\ud y \ud^2\kt \ud^2S_\perp}
\right)
\left/
\left(
\frac{\ud N_{pp}}{\ud y \ud^2\kt\ud^2S_\perp}
\right)
\right.
. 
\ee
The experimentally measured quantity is naturally 
integrated over the transverse area of the collision system.
In the case of a symmetric collision \eq\nr{eq:RAB} reduces to 
$\raa$ and for a pA collision to $\rpa$.
In the MV model the physical interpretation of the saturation scale
is straightforward as resulting from an incoherent sum of the color charges of
high-$x$ partons. This leads to the scaling of the color charge density as
 $\mu_\subA^2=A^{1/3} \mu_p^2$, and the normalization factor 
in $\raa$ reduces to $A^{2/3}$ which is simply the number of 
binary collisions per unit area $\ud N_\textrm{coll}/\ud^2S_\perp$.
In the proton-nucleus case
the ratio of the spectra in \eq\nr{eq:RAB} is again normalized by
the number of collisions $N_\textrm{coll}$, which now  scales like $A^{1/3}$.

\begin{figure}
\includegraphics[width=0.48\textwidth]{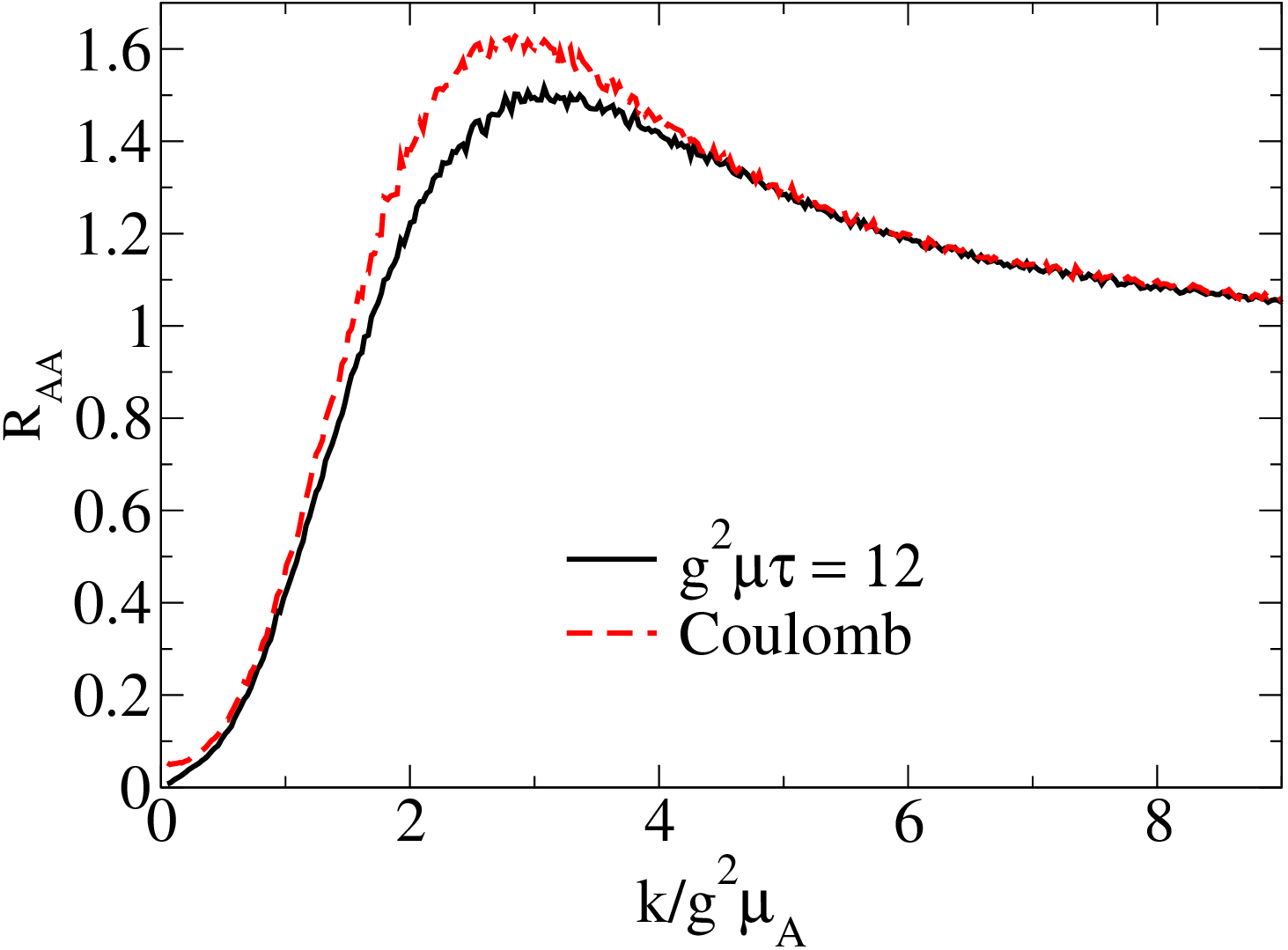}
\caption{
 Nuclear modification factor $\raa$, normalized with $(g^2\mu)^4$
 of the gluon spectrum
for $g^2\mu_\subA = 2\gev$, $g^2\mu_p = 0.32\gev$ and $N_y=500$.
}
\label{fig:RAA}
\end{figure}

Our numerical results for $\rpa$ and $\raa$
are shown in \figs\ref{fig:RpA} and \fig\ref{fig:RAA} respectively.
  They both exhibit a Cronin enhancement, 
peaked respectively at  about two and three times the 
saturation scale $g^2\mu_\subA$. The Cronin peak has been 
indeed observed in ``pA'' collisions, whereas in the ``AA'' 
case a large suppression of a factor 5 has been observed at RHIC,
commonly understood in terms of energy loss in the
hot and dense medium produced in  the
 collision. The CYM calculation does not account for such a 
final sate effect.

\begin{figure}
\includegraphics[width=0.48\textwidth]{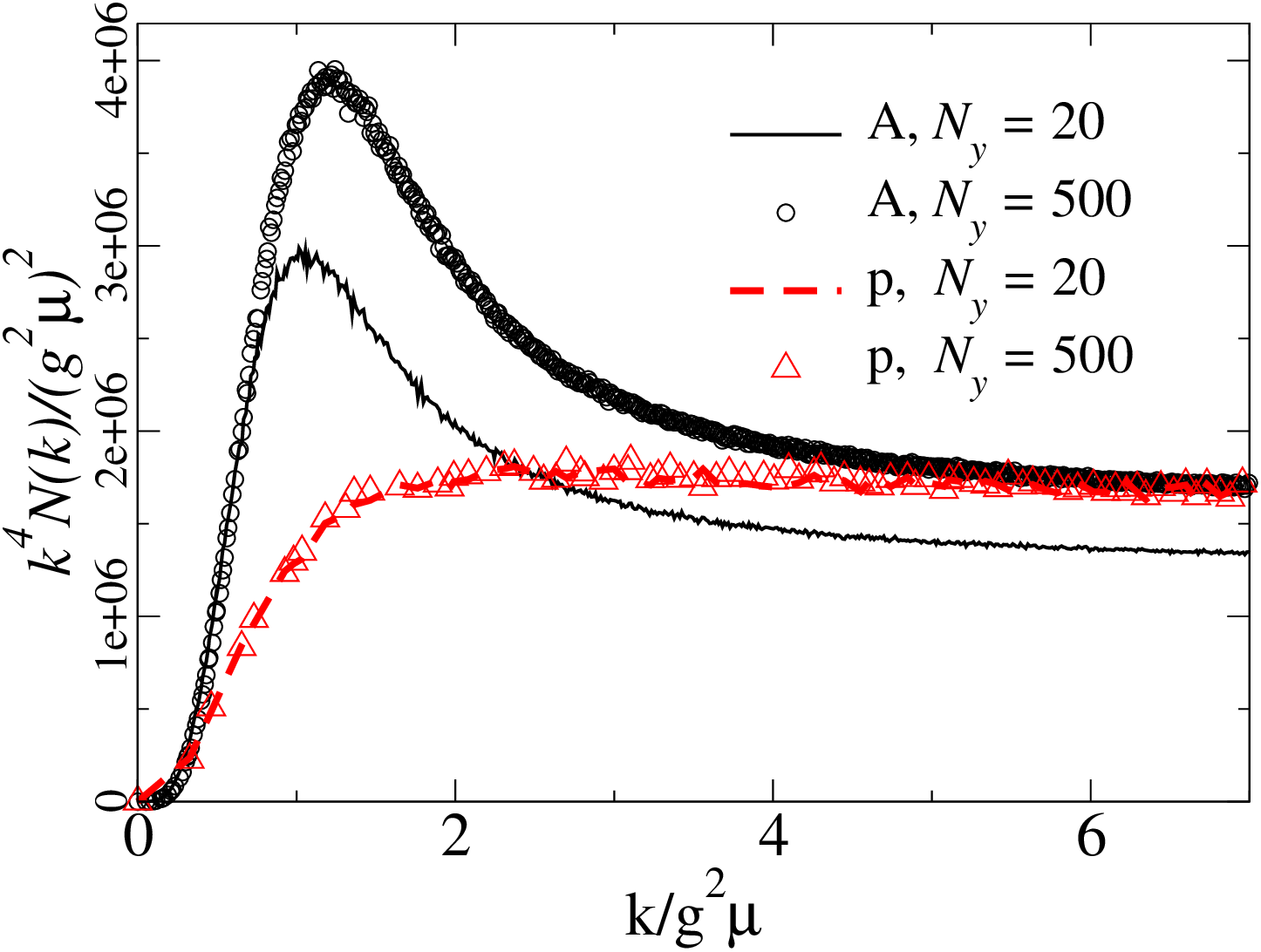}
\caption{
Scaled fundamental representation Wilson line correlators 
$(\ktt^4/ (g^2\mu)^2) \cN(\kt/g^2\mu)$ 
for $g^2 \mu_\subA = 2\gev$, $g^2\mu_p = 0.32\gev$, $m=0.1\gev$. 
Results are shown for
both  $N_y=20$ and $N_y =500$; the latter ones are obtained
from the configurations used for 
\figs\ref{fig:RpA} and~\ref{fig:RAA}.
The scaling $\cN(\kt) \sim (g^2 \mu)^2/\ktt^4$ is achieved only for 
the larger value of $N_y$. When the horizontal axis is scaled with the position
of the peak (i.e. $\qs$) instead of $g^2\mu$, 
the difference between the different values
of $N_y$ is smaller.
}
\label{fig:wlines}
\end{figure}

There is an additional remark  we must make concerning the numerical
calculation.  In order for the nuclear modification factor to approach
one for large transverse momenta, the unintegrated gluon distribution
(or Wilson line correlator $\cNt(\kt)$, see \eq\nr{eq:ft} below) must
approach the asymptotic behavior $\sim (g^2\mu)/\kt^4$ with a constant
of proportionality that is independent of any infrared scale in the
problem. 
We have numerically found (see Ref.~\cite{Lappi:2009xa} for more
details) that
to achieve this one must discretize the longitudinal
coordinate in constructing the Wilson line on a very fine grid. 
 This means that to
approach the right large $\kt$ limit in nuclear modification factor
one must use a much larger value of $N_y$ than was needed for the
determination of $\qs$ in terms of $g^2\mu$~\cite{Lappi:2007ku} or for
the single or double inclusive gluon spectra in the bulk region of
momenta around $\qs$~\cite{Lappi:2009xa}.  The slow convergence
in the $N_y\to \infty$ limit is demonstrated in \fig\ref{fig:wlines}
for $\cN$ in the fundamental representation.

As we already saw in the previous section, in the dilute-dense limit
the approximation of linearized final state evolution gives the correct result.
For the nuclear modification ratio $\rpa$ this is demonstrated in 
\fig\ref{fig:RpA}, where this ratio is plotted using both the full CYM result
and the Coulomb gauge approximation. A slight difference between the full result and 
the Coulomb gauge approximation for $\raa$ is seen in 
\fig\ref{fig:RAA}, the Coulomb gauge approximation leading to a slight 
overestimate.

\section{The $\ktt$-factorized approximation}
\label{sec:ktfact}

\begin{figure}
\includegraphics[width=0.48\textwidth]{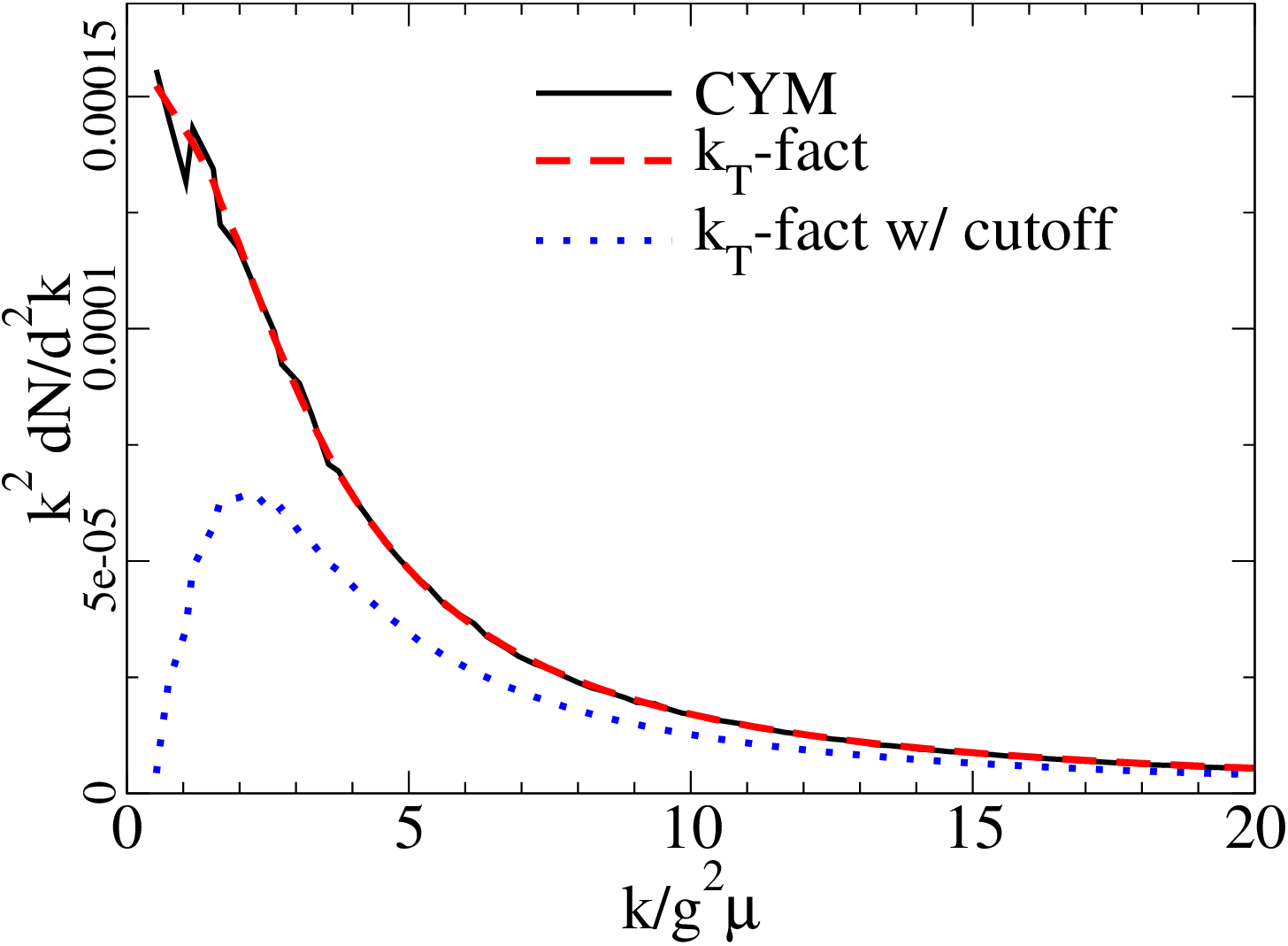}
\caption{$\ktt$-factorized results compared to the full CYM calculation in the 
dilute limit
$g^2\mu_\subA = 0.2\gev$, $m=0.1\gev$.}
\label{fig:ktfactdilute}
\end{figure}

As is discussed in detail in Ref.~\cite{Blaizot:2008yb}, there is no valid 
$\ktt$-factorized expression for the gluon multiplicity in the fully nonlinear
case of AA-collisions, in contrast to the case of
the dilute ``pA'' limit.
Different $\ktt$-factorized approximations for nucleus-nucleus collisions have 
nevertheless been extensively used in the literature 
and it is therefore instructive to check how good these approximations are.

Our starting point is the following $\ktt$-factorized ansatz for the gluon multiplicity
\begin{multline}\label{eq:ktfact}
\frac{\ud N}{\ud y \ud^2\kt}
= \frac{\pi \ra^2}{(2\pi)^2} \frac{\nc^2-1}{2\nc} \frac{2}{\pi g^2 \kt^2} 
\int \frac{\ud^2\qt}{(2\pi)^2} 
\left[\qt^2 \cNt_{\subA}(\qt)\right]
\\ \times
 \left[(\kt-\qt)^2 \cNt_{\subB}(\kt-\qt)\right],
\end{multline}
where (for nucleus $A$)
\begin{equation}\label{eq:ft}
 \cNt_{\subA}(\qt) = \frac{1}{\nc^2-1} \int \! \ud^2 \rt \; e^{i\rt\cdot \qt}
\left\langle \tr \widetilde{U}^\dag(\xt+\rt) \widetilde{U}(\xt) \right\rangle
\end{equation}
and similarly for $ \cNt_{\subB}$ with $U$ replaced by $V$.
The tilde refers to the adjoint representation, since the adjoint correlator
is what appears in the A-side of the pA $\ktt$-factorization formula. We emphasize 
that there is no derivation of \eq\nr{eq:ktfact} in the ``AA''-case, it is an ansatz 
that a) is symmetric in the two nuclei and b) reduces to the correct known
result in  both the ``pA'' and the ``Ap'' limits, when one of the sources
is taken to be weak.

The formula \nr{eq:ktfact} lends itself to simplifications in different limiting
cases.
In particular, in the large momentum limit $|\kt|\gg \qs$
the integral is dominated by the regions $\qt \sim 0$ and $\qt \sim \kt$. 
In the first one of these regions one can approximate $\kt-\qt \approx \kt$
in the second factor and pull it outside the integral; in the second region the same
can be done to the other factor. This  leaves the result
\begin{multline}\label{eq:ktfactlargept}
\frac{\ud N}{\ud y \ud^2\kt}
= \frac{\pi \ra^2}{(2\pi)^2} \frac{\nc^2-1}{2\nc} \frac{2}{\pi g^2}
\\
\left[
 \cNt_\subA(\kt) xG_\subB(x,\kt^2)
+ 
 \cNt_\subB(\kt) xG_\subA(x,\kt^2)
\right]
\end{multline}
that is proportional to the dipole cross section or Wilson line correlation function
$\cNt(\kt)$ in one nucleus and the integrated gluon distribution
\begin{equation}
xG(x,\kt^2) = \int^{|\kt|} \frac{\ud^2\qt}{(2\pi)^2} \qt^2 \cNt(\qt)
\end{equation}
in the other one. 
At $\ktt \gg \qs$ we have parametrically $\cNt(\kt)\sim \qs^{2}/\kt^4$ 
(note that $\cNt(\xt)$ is dimensionless)
and $xG \sim \qs^2$.

\begin{figure}
\includegraphics[width=0.48\textwidth]{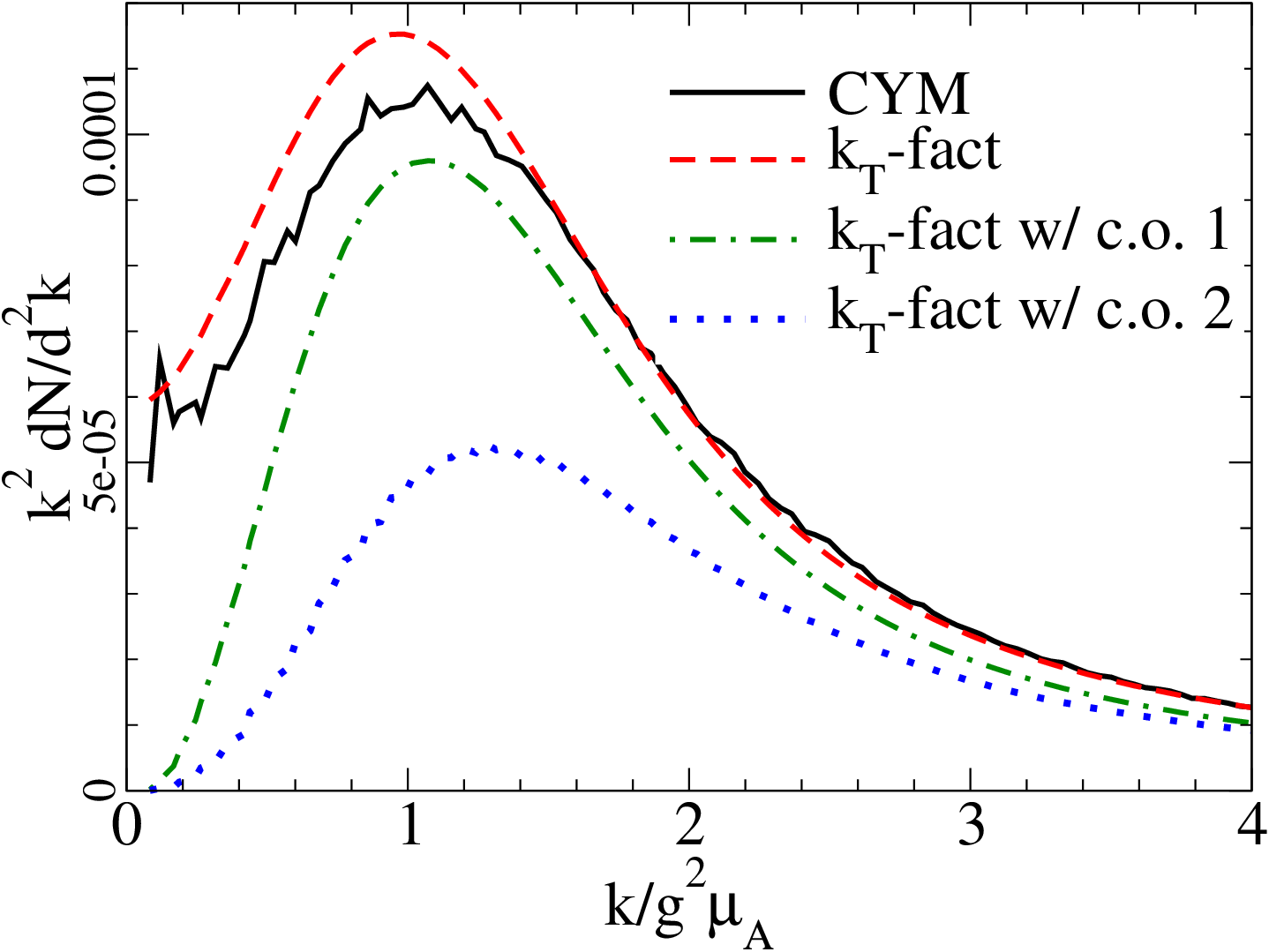}
\caption{$\ktt$-factorized results compared to the full CYM calculation for the 
``pA'' case 
$g^2\mu_\subA = 1.25 \gev$,  $g^2\mu_p = 0.008 \gev$. 
The dot-dashed curve (``c.o. 1'' i.e. ``cutoff 1'') 
is the case where the momentum from the
proton is restricted to be less than that of the produced gluon (i.e. 
nucleus $A$ in \eq\nr{eq:ktfact} is taken as the proton and there is a cutoff
$\theta(|\kt|-|\qt|)$). In the
dotted curve (``c.o. 2'') the cutoff is applied to the momentum 
from the nucleus (i.e. $B$ in  \eq\nr{eq:ktfact} is the proton and the
cutoff is still $\theta(|\kt|-|\qt|)$). 
The full numerical result shows a significant oscillation 
which is typical for a calculation at a finite ending time $g^2\mu_\subA\tau=60$.
$N_y=20$ and $m=0$.}
\label{fig:ktfactpA}
\end{figure}

Two main variants of the $\ktt$-factorized formula \eq\nr{eq:ktfact}
have also been used in the literature. 
\begin{itemize}
\item The $\qt$-integration is often cut off (see
  e.g.~\cite{Armesto:2004ud,Hirano:2004rs,Drescher:2006pi,Drescher:2006ca,Albacete:2007sm})
  with $\theta(|\kt|-|\qt|)$; this is asymmetric in the two nuclei but
  does have the advantage of making the spectrum IR-finite, which it
  is otherwise not. This cutoff is relatively easy to implement in our
  numerical evaluation, and we shall discuss its effect below.
\item Instead of $\qt^2 \cNt(\qt)$ (which we know from the pA-case to
  be the object appearing on the A-side) one sometimes replaces the
  $\qt^2$ by a $1/\rt^2$ inside the Fourier-transform \eq\nr{eq:ft}.
  This gives an unintegrated gluon distribution $\varphi(\qt)$ 
  (the ``WW'' distribution,
  see~\cite{Kharzeev:2003wz,Albacete:2003iq,Blaizot:2004wu,Gelis:2006tb}
  ) that is related to the number of gluons as defined in LC
  quantization and behaves like $\ln |\qt|$ for small momenta. This is
  closer to the KLN ansatz~\cite{Kharzeev:2000ph,Kharzeev:2001gp}
  $\varphi(\qt) \sim \textrm{cst.}/\as$.  Due to the logarithmic
  divergence of the unintegrated gluon distribution at small $\qt$
  this approximation is more difficult to treat directly in our
  numerical setup and we do not study it further here.
\end{itemize}

\begin{figure}
\includegraphics[width=0.48\textwidth]{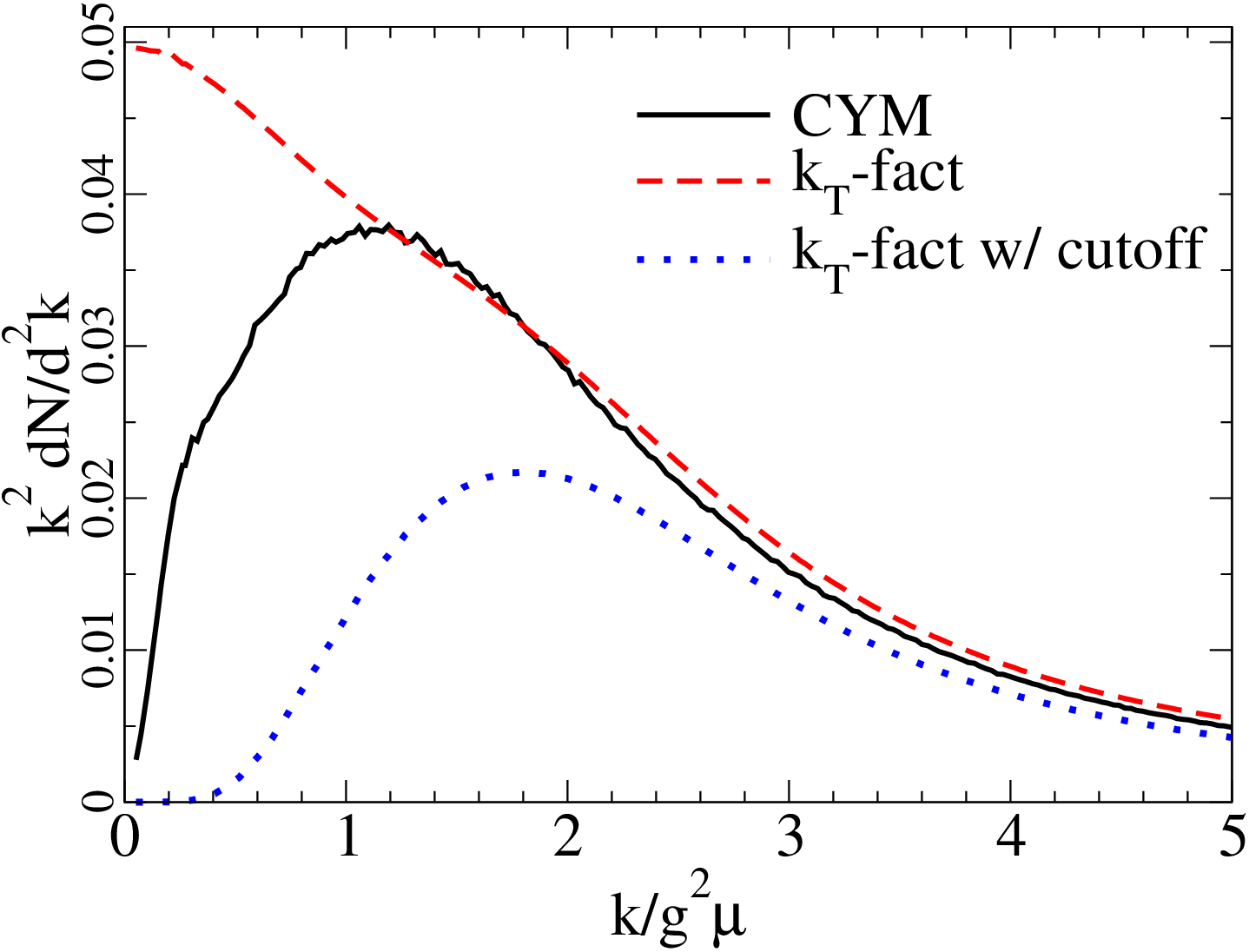}
\caption{$\ktt$-factorized results compared to the full CYM calculation 
in the saturated  strong field regime, with 
$g^2\mu = 2\gev$, $m=0.1\gev$ $N_y=20$ (same configurations as in 
\fig\ref{fig:lccoul}).}
\label{fig:ktfactdense}
\end{figure}

Figure \ref{fig:ktfactdilute} shows the comparison of \eq\nr{eq:ktfact} with 
or without the cutoff $\theta(|\kt|-|\qt|)$ to
the full numerical result in the dilute regime where one expects agreement.
One sees that indeed, as can be shown analytically, in the dilute case
the  $\ktt$-factorized expression (without the cutoff) is accurate, but 
the cutoff changes this already at quite high momenta.
 Figure \ref{fig:ktfactpA} shows the comparison between 
\eq\nr{eq:ktfact} with and without the cutoff 
and the full result in the asymmetric ``pA'' case. Again we confirm the
analytical calculation showing that the  $\ktt$-factorized approximation
is good also in the dense-dilute case.
Note the dependence on
whether the momentum that is cut off is that of the proton or the nucleus.
Figure~\ref{fig:ktfactdense} shows the same comparison 
between \eq\nr{eq:ktfact} and the full CYM result for the ``AA'' case.
We see that while the cutoff does make the
spectrum IR finite, it does so at the expense of deviating from
the full result already at high momenta, $\ktt \lesssim 2g^2\mu$.
Unlike in the previous dilute cases, also 
the result without the cutoff deviates from the full result
for $\ktt \lesssim g^2\mu$. Comparing \figs\ref{fig:ktfactdense} 
and~\ref{fig:lccoul} one immediately sees that the approximation 
using the Coulomb gauge fields at $\tau=0$ is much more
accurate than the $\ktt$-factorized one.

Note that our result on  the form of
the spectrum in $\ktt$-factorization does not invalidate computations where
$\ktt$-factorization has been used to study the dependence
of the integrated multiplicity on energy, rapidity, centrality etc.
The integrated multiplicity will still be, by dimensional reasons,
proportional to $\qs^2$. Thus the determining aspect of these phenomenological
applications is the dependence of $\qs$ on impact parameter and $x$, not
the precise shape of the initial gluon spectrum, which will be modified
later in the plasma phase.

\section{Conclusion and perspectives}

In conclusion, we have studied numerically the spectrum of gluons in the
Glasma fields in the initial stages of a heavy ion collision. We have 
compared the results obtained by approximations where one includes
the nonlinear interactions of the gluon fields only in the initial condition
(at $\tau=0$) to the full numerical CYM calculation. The separation 
between initial and final state effects is not a gauge invariant one,
and thus our discussion naturally involves finding a gauge that minimizes
the final state effects. We
find that in the Fock-Schwinger + transverse Coulomb gauge the effect
of final state rescatterings on the spectrum is surprisingly small except 
at very small momenta.
It would be interesting to see how the corrections to the linearized approximation 
converge towards the full CYM result.

We have also compared our  results to those obtained assuming 
$\ktt$-factorization. While 
in the ``pp'' and ``pA'' cases the results are the same, as is well known,
in the fully nonlinear ``AA'' case  $\ktt$-factorization
gives a poorer description of the gluon spectrum
for $\ktt \lesssim \qs$ than the linear approximation used in this paper,
with a marked sensitivity to the infrared cutoff.

\acknowledgments{
We acknowledge numerous discussions with F. Gelis and R. Venugopalan on this work
and related topics. T.L. is supported by the Academy of Finland, project 126604.
}

\appendix

\section{Proton-Nucleus collisions and dilute limit}

It was shown in Ref.~\cite{Blaizot:2008yb} that  \eq\nr{eq:spectrum2} evaluated
with the $\Omega=VU$-gauge fields gives the known $\ktt$-factorized
formula for the gluon spectrum in the ``pA'' case. 
It was shown in 
Ref.~\cite{Dumitru:2001ux} that a Coulomb gauge calculation of the gluon 
spectrum  in pA collisions gives the same result. Let us here briefly show how 
this comes about evaluating our \eq\nr{eq:spectrum2} in Coulomb gauge.
Assuming nucleus B to be a proton one can expand the gauge field to first 
order in $\rho_\subB$, and we get
\beq\label{eq:expansion1}
\Acoul^\eta_{(1)}|_{\tau=0^+}&=& \frac{ig}{2}\Omega_{(0)}[A_{\subA}^i,A_{\subB (1)}^i]\Omega^\dag_{(0)}\\
\Acoul_{(1)}^i |_{\tau=0^+}&=& \Omega_{(1)}A^i_\subA\,\Omega^\dag_{(0)}+\Omega_{(0)} A^i_\subA\,\Omega^\dag_{(1)}+\Omega_{(0)} A^i_{\subB(1)}\,\Omega^\dag_{(0)}\nn 
&&-\frac{1}{ig} \Omega_{(1)}\del^i\Omega^\dag_{(0)}-\frac{1}{ig} \Omega_{(0)}\del^i\Omega^\dag_{(1)},\nn
\eeq
where $\Omega_{(0)}\equiv U$, 
\be
A_{\subB(1)}^i=-\frac{\del^i}{\nabt^2\del^+}\rho_\subB,
\ee
and $\Omega_{(1)}$ can be extracted from Eq. (\ref{eq:gaugelink}), 
\be
\Omega^\dag_{(1)}\equiv igU^\dag \frac{\del^i}{\nabt^2}\left(U A^i_{\subB(1)}U^\dag\right).
\ee
We obtain after some algebra 
\beq\label{eq:expansion2}
\left. \Acoul^\eta_{(1)}\right|_{\tau=0^+}
\!\!\!\!\!
&=&-(\del^i U)A_{\subB (1)}^iU^\dag-U A_{\subB (1)}^i (\del^i U^\dag), \\
\left. \Acoul_{(1)}^i \right|_{\tau=0^+}
\!\!\!\!\!
&=& \left(\delta^{ij}-\frac{\del^i\del^j}{\nabt^2}\right) \left(U A_{\subB (1)}^i  U^\dag\right).
\eeq
Plugging \eq\nr{eq:expansion2} into \nr{eq:spectrum2}
leads to the well known $\ktt$-factorization formula for gluon 
production in proton-nucleus collisions. \\

\section{Hamiltonian variables used in the numerical calculation}

\begin{figure}
\includegraphics[width=0.48\textwidth]{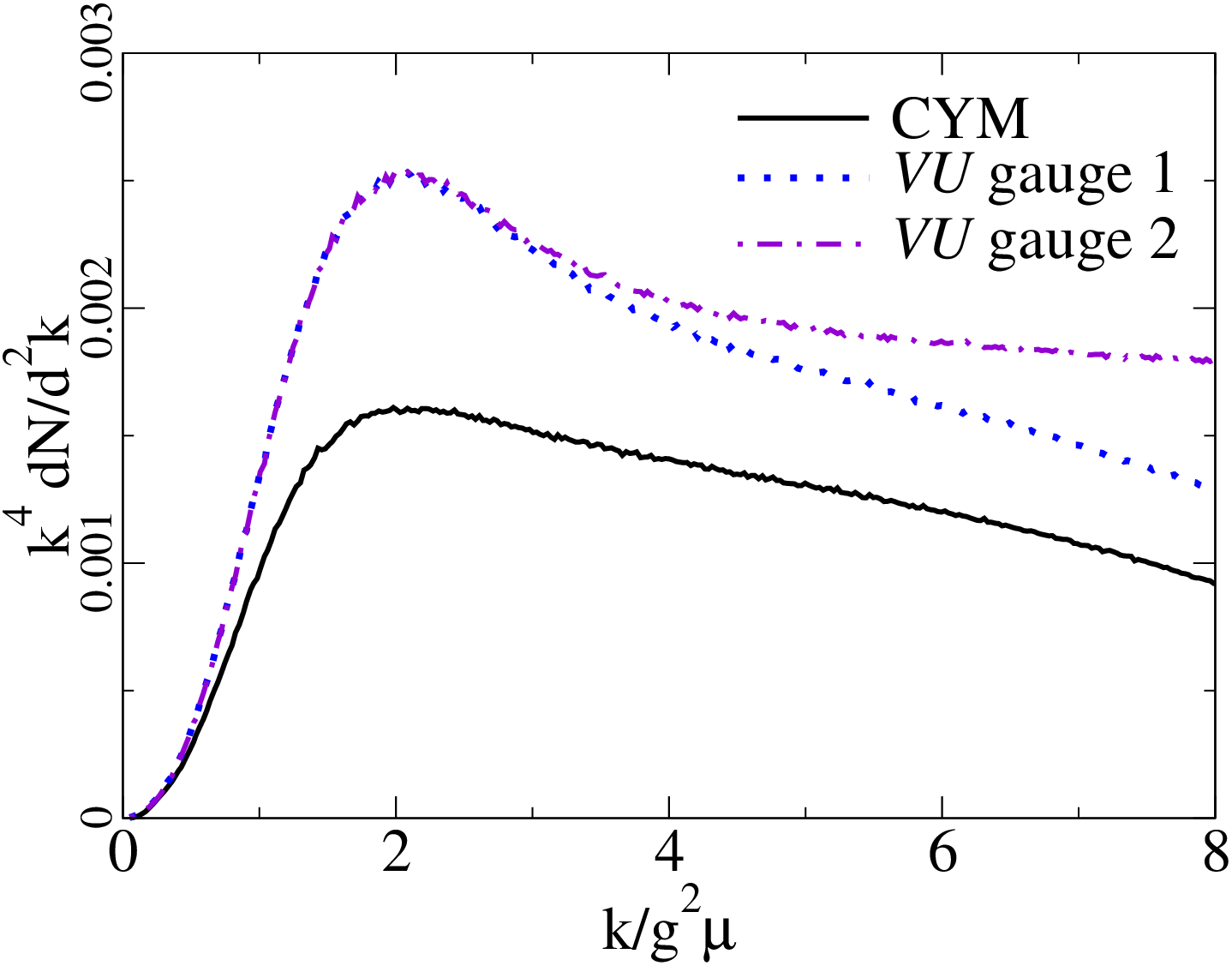}
\caption{Different discretization methods of the VU-gauge, see text for the 
explanation of the labels. $g^2\mu=2\gev$, $N_y=1$ $m=0$.
 Note that the spectrum is multiplied by $\ktt^4$ to show the effects at large $\ktt$.}
\label{fig:bmtdiscr}
\end{figure}

\begin{figure}
\includegraphics[width=0.48\textwidth]{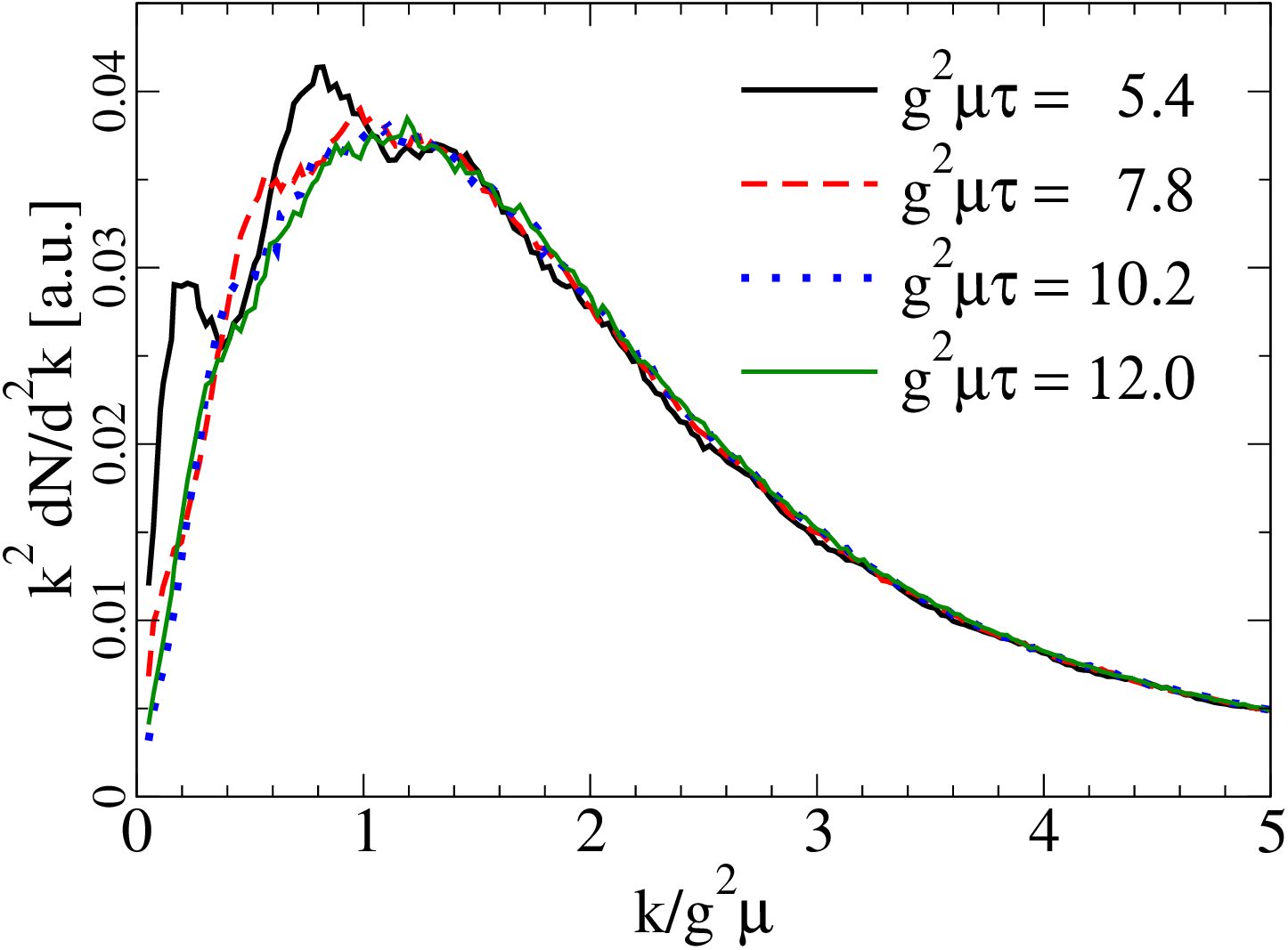}
\caption{CYM gluon spectrum at different times. 
$N_y=20, m=0.1\gev$ (same configurations as in \fig\ref{fig:lccoul}).}
\label{fig:cymtime}
\end{figure}

In the numerical calculations it is customary to work in a Hamiltonian 
formalism  with the gauge potentials and  electric fields
\begin{eqnarray}\label{eq:hamvariables}
E^i &=& \tau \dot{A}_i
\\
A_\eta &=& -\tau^2 A^\eta = x^+ A^- - x^- A^+ 
\\ 
E^\eta &=& \frac{1}{\tau} \dot{A}_\eta
\end{eqnarray}
The transverse gauge potential is represented in terms
of the link matrix
\begin{equation}
U_i = e^{-iga A_i}.
\end{equation}
In Refs.\cite{Krasnitz:1998ns,Lappi:2003bi} the notation $A_\eta
\equiv \phi$ was used, while the longitudinal electric field was
denoted $E^\eta = \pi$ in \cite{Lappi:2003bi} and $E^\eta = p$ in
\cite{Krasnitz:1998ns}.  In terms of these Hamiltonian variables the
initial condition \eq\nr{eq:initcond} for the longitudinal field is
\begin{eqnarray}\label{eq:haminitcond}
A_\eta|_{\tau=0^+} &=& 0
\\
E^\eta|_{\tau=0^+} &=& - ig [A^i_{(1)},A^i_{(2)}].
\end{eqnarray}

In terms of the Hamiltonian variables the expression for
the multiplicity with linearized final state evolution,
\eq\nr{eq:spectrum2}, reads
\begin{multline}\label{eq:bmtmultitau0}
\frac{\ud N}{\ud y \ud^2 \kt} = \frac{1}{(2 \pi)^2} \frac{1}{\pi \kt^2}
\Big\langle
\left|\kt \times {\bf\cal A}(\kt) \right|^2
\\
+ \left|E^\eta (\kt) \right|^2
 \Big\rangle_{\tau=0^+}
\end{multline}
The relation between the notations in Ref.~\cite{Lappi:2003bi}
and Ref.~\cite{Blaizot:2008yb} is, 
with  \cite{Lappi:2003bi} on the left and \cite{Blaizot:2008yb}
on the right of the equal signs,
$U_{(1)} = V^\dag$, $U_{(2)} = U^\dag$, $ A^i_{(1)} = - A^i_\subB$ and
$ A^i_{(2)} = - A^i_\subA$ (the sign is compensated by the opposite sign in the covariant 
derivative $\partial^\mu + i g A^\mu_{(m)} = \partial^\mu - i g A^\mu_{\subA,\subB}$).
Between these two references the Wilson line is exchanged with its
Hermitian conjugate (including changing the direction
of the path ordering in the path ordered 
exponential);  in the conventions of Ref.~\cite{Lappi:2003bi} the pure gauge field is
\begin{equation}
 A^i_{(1)} = -\frac{i}{g} U_{(1)} \partial^i  U_{(1)}^\dag.
\end{equation}

\section{Discretization of \eq\nr{eq:spectrum2} in the VU gauge}

The formulas we need to discretize are \eq\nr{eq:lcinitc1}:
\begin{equation}\label{eq:alfa}
\Acoul^i  = V \left( 
 U \B U^\dag- \B \right) V^\dag,
\end{equation}
and \eq\nr{eq:lcinitc2}
\begin{multline}\label{eq:beta1}
\Acoul^\eta
V \left[\left( \partial^i  U \right) \B^i U^\dag
+  U \B^i \left( \partial^i U^\dag \right) \right] V^\dag,
\end{multline}
which can also be written as
\begin{equation}\label{eq:beta2}
\Acoul^\eta  = t^a V_{ab} \left(\partial^i U_{bc}\right) 
\B^{i,c}.
\end{equation}

A straightforward way to discretize \eq\nr{eq:alfa} is to define
\begin{multline}\label{eq:defalfalat}
\Acoul^i (\xt)
=  \half \bigg[ V(\xt + \jt)
 \Big( U(\xt+\jt) \B^j(\xt) U^\dag(\xt+\jt) 
\\
- \B^j(\xt)\Big) U^\dag(\xt+\jt)
\\
+
 V(\xt ) \Big( U(\xt) \B^j(\xt) U^\dag(\xt) 
- \B^j(\xt)\Big) V^\dag(\xt)
\bigg],
\end{multline}
where $\jt$ is a vector of length $a$ in the $j$-direction.
Here the lattice gauge field is really the antihermitian part of the link matrix:
\begin{equation}\label{eq:latA}
{\B}_i(\xt) = -{\B}^i(\xt) = \frac{1}{i 2 g a }\left( U_{\subB,i}(\xt)-U^\dag_{\subB,i}(\xt) \right),
\end{equation}
where the link matrices representing the transverse pure gauge fields of
the individual nuclei are
\begin{eqnarray}
U^{\subA}_{i}(\xt) &=& U^\dag(\xt)U(\xt+ \itt) \\ \nonumber
U^{\subB}_{i}(\xt) &=& V^\dag(\xt)V(\xt+ \itt).
\label{eq:aher}
\end{eqnarray}
In terms of this we then write down the discretized derivative needed in 
\eq\nr{eq:spectrum2} as
\begin{equation}\label{eq:alfader}
\epsilon^{ij} \partial_i  \Acoul^j (\xt)
= 
\epsilon^{ij}\left[
\Acoul^j (\xt+\itt) - \Acoul^j (\xt)
\right].
\end{equation}

The longitudinal field can is then discretized in the same spirit as
the transverse one. 
The two versions \eqs\nr{eq:beta1} and~\nr{eq:beta2} lead to 
different-looking discretizations
\begin{widetext}
\begin{multline}\label{eq:betadisc1}
\Acoul^\eta =
\half V(\xt+\itt)
\bigg[U(\xt+\itt) {\B}_i(\xt)  U^\dag(\xt+\itt)
-  U(\xt) {\B}_i(\xt)  U^\dag(\xt)
\bigg]V^\dag(\xt+\itt)
\\
+
\half
V(\xt)\bigg[
U(\xt+\itt) {\B}_i(\xt)  U^\dag(\xt+\itt)
-  U(\xt) {\B}_i(\xt)  U^\dag(\xt)
\bigg]V^\dag(\xt)
\end{multline}
and
\begin{multline}\label{eq:betadisc2}
\Acoul^\eta =
\half
V(\xt+\itt)\bigg[ U(\xt+\itt) {\B}_i(\xt)  U^\dag(\xt)
+  U(\xt) {\B}_i(\xt)  U^\dag(\xt+\itt)
- 2  U(\xt) {\B}_i(\xt)  U^\dag(\xt)
\bigg]V^\dag(\xt+\itt)
\\
+
\half
V(\xt)\bigg[ U(\xt+\itt) {\B}_i(\xt)  U^\dag(\xt)
+  U(\xt) {\B}_i(\xt)  U^\dag(\xt+\itt)
- 2  U(\xt) {\B}_i(\xt)  U^\dag(\xt)
\bigg]V^\dag(\xt).
\end{multline}
\end{widetext}
Of these two we prefer the first one \eq\nr{eq:betadisc1} because of its
 relative simplicity. The combination \eqs\nr{eq:defalfalat} and 
\nr{eq:betadisc1} is labeled as ``$VU$-gauge 2'' in \fig\ref{fig:bmtdiscr}.

An alternative discretization method is suggested by the observation that the 
version of the light cone gauge proposed in \cite{Blaizot:2008yb} is equivalent
to taking the gauge fields in the Fock-Schwinger gauge and performing a gauge rotation
with the product of the Wilson lines of the two nuclei.
The lattice version of initial conditions for the transverse fields is 
obtained \cite{Krasnitz:1998ns,Lappi:2003bi} by 
solving the link matrix $U_i(\xt)$ from the equation 
\begin{equation}\label{eq:trlatinitcond}
\tr \left[t_a \left(\left(U^{\subA}_i +U^{\subB}_i\right)
\left(1+U_i^\dag\right) - \hc \right) \right] =0.
\end{equation}
Here the pure gauge link matrices $U_i^{A,B}$  (see 
\eq\nr{eq:latA}) correspond to the pure
gauge fields of the two nuclei separately.
For SU(2) this equation can be solved in closed
form, but in the case of SU(3) that we are interested in here it is
solved numerically by an iterative procedure.  This link matrix $U_i$ is
then used in the initial condition for the longitudinal field
\begin{multline}\label{eq:llatinitcond}
E^\eta(\xt) =  \frac{-i}{4g} \sum_{i} \bigg[
\Big(U_i(\xt) - 1\Big)
\Big(U_i^{\dag \subA}(\xt)-U_i^{\dag \subB}(\xt) \Big)
\\
+
\Big(U_i^{\dag}(\xt-\itt) - 1\Big)
\Big(U_i^{\subA}(\xt-\itt)-U_i^{\subB}(\xt-\itt) \Big) -\hc
\bigg].
\end{multline}
The fields obtained from \eqs\nr{eq:trlatinitcond} and \nr{eq:llatinitcond}
are then gauge transformed with either $\Omega = UV$
or $\Omega = VU$ and taking the antihermitian part of the link matrix
as the gauge field in \eq\nr{eq:spectrum2}.
$VU$-gauge 1 in \fig\ref{fig:bmtdiscr}. We emphasize that the small difference 
between these two methods only appears at momenta of the order of the lattice UV 
cutoff. We observe that the UV behavior of the latter method is closer
to the Coulomb gauge curve, and it is the one we use in the rest of this paper.

\section{Multiplicity in the CYM calculation}
\label{sec:cymmulti}

In the CYM computations one has to choose some finite time $\tau$ 
at which to perform the Fourier
decomposition of the fields for computing the spectrum. In the boost invariant 
calculations the interactions of the fields become weaker with time, so the result
does not depend very strongly on $\tau$ for $\tau \gg 1/\qs$. The remaining residual 
dependence is demonstrated in \fig\ref{fig:cymtime}, which shows the resulting gluon 
spectrum for $g^2\mu\tau = 4.2, 7.8$ and $12$.

\begin{figure}
\includegraphics[width=0.48\textwidth]{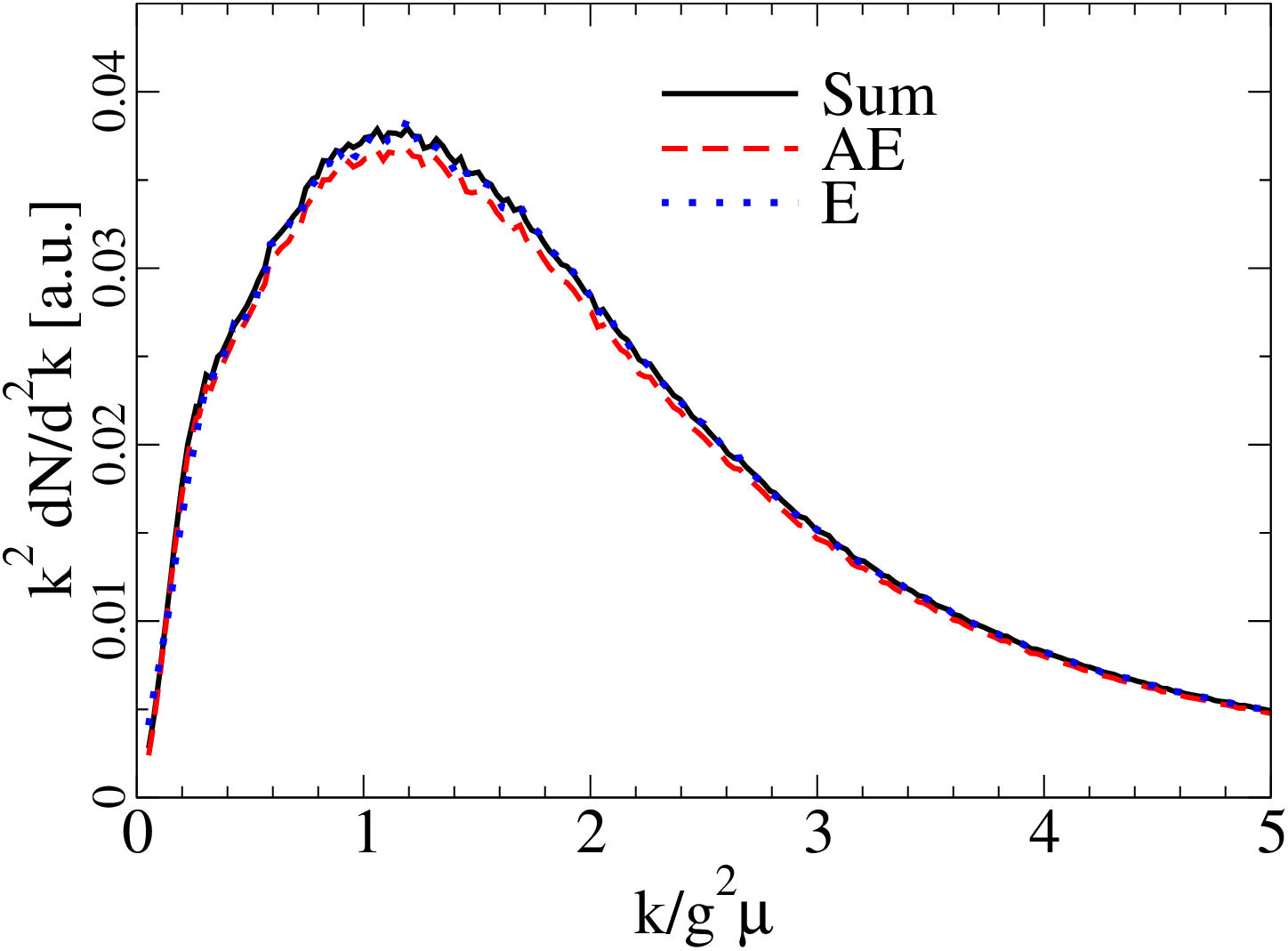}
\caption{CYM gluon spectrum at $g^2\mu\tau=12$ using different methods. The solid 
line labeled ``Sum'' is the one defined by \eq\nr{eq:multisum}
that we use, unless otherwise stated, in this paper. The dashed line
(``AE'') is the formula \nr{eq:multiae} that cancels the unknown dispersion
relation and the dotted one (``EE'') the one of \eq\nr{eq:multiee}.
$N_y=20, m=0.1\gev$ (the field configurations are the same as in \fig\ref{fig:lccoul}).}
\label{fig:cymmethod}
\end{figure}

Let us then turn to the question of defining the multiplicity corresponding to a 
classical field configuration that does not yet evolve completely linearly, as 
is the case in practical numerical computations. The straightforward method used
in \cite{Lappi:2003bi} is to start from the electric field part of the Hamiltonian 
in Coulomb gauge. The electric fields represent approximately half of the total energy
of the system (which is easily verified in the full numerical computation) and because
the electric field part of the Hamiltonian is quadratic in the canonical momenta
one can obtain a decomposition of the total energy of the system into 
transverse momentum modes. If we now \emph{assume} that each of these modes has a free
dispersion relation $\omega(\qt)=|\qt|$ (or the corresponding lattice equivalent in 
numerical computations), we get our first definition of the multiplicity
\begin{multline}\label{eq:multiee}
\frac{\ud N}{\ud y \ud^2\kt}= \frac{1}{(2 \pi)^2}\frac{1}{|\kt|} \Big[
\frac{1}{\tau} E^i_a(\kt)E^i_a(-\kt) 
\\+\tau \pi_a(\kt) \pi_a(-\kt)
\Big].
\end{multline}

The dispersion relation of the interacting theory is, however, not free. The implications
of this observation for our situation were observed already in the first CYM determination
of the Glasma multiplicity \cite{Krasnitz:2000gz}. By looking at correlators
of the fields $A_\mu$ and the corresponding electrical fields separately one 
can numerically determine this dispersion relation.  It was observed in 
\cite{Krasnitz:2000gz} that this dispersion relation exhibits a mass gap that 
decreases with time as $m^2 \sim g^2\mu/\tau$.
By taking products of the correlators one can then
construct a formula for the multiplicity where the dispersion relation cancels out:
\begin{widetext}
\begin{multline}\label{eq:multiae}
\frac{\ud N}{\ud y \ud^2\kt}= \frac{1}{(2 \pi)^2}
\sqrt{
\frac{1}{\tau} E^i_a(\kt)E^i_a(-\kt) +\tau \pi_a(\kt) \pi_a(-\kt)
}
\sqrt{
\tau A^i_a(\kt)A^i_a(-\kt) +\frac{1}{\tau} \phi_a(\kt) \phi_a(-\kt)
}.
\end{multline}
\end{widetext}

The third option, and the one we will use unless otherwise stated, is based 
on the formal derivation from a reduction formula~\cite{Gelis:2007kn}.
Again this  is similar in spirit to our first definition and 
amounts to assuming a free dispersion relation $\omega(\kt)$
and taking the sum  of the canonical fields and momenta
as 
\begin{widetext}
\begin{multline}\label{eq:multisum}
\frac{\ud N}{\ud y \ud^2\kt}= 
\frac{1}{2 (2 \pi)^2}
\bigg\{
\frac{1}{\omega(\kt)}
\bigg[
\frac{1}{\tau} E^i_a(\kt)E^i_a(-\kt) 
+\tau \pi_a(\kt) \pi_a(-\kt) \bigg]
+
\omega(\kt)
\bigg[
\tau A^i_a(\kt)A^i_a(-\kt) +\frac{1}{\tau} \phi_a(\kt) \phi_a(-\kt)
\bigg]
\bigg\}.
\end{multline}
\end{widetext}
When done properly the reduction formula also gives an additional
(small in practice) contribution that is antisymmetric in $\kt
\leftrightarrow -\kt$. This contribution vanishes when the single
inclusive multiplicity is averaged over configurations and already for
a single configuration when the spectrum is averaged over the
azimuthal angle of $\kt$. It is, however,
important for multigluon correlations in the
Glasma~\cite{Gelis:2008ad,Gelis:2008sz,Lappi:2009xa,Dusling:2009ni}. The
agreement between the three methods is illustrated in
\fig\ref{fig:cymmethod}.

\bibliography{spires}

\providecommand{\href}[2]{#2}\begingroup\raggedright\begin{thebibliography}{10}

\bibitem{Iancu:2003xm}
E.~Iancu and R.~Venugopalan in {\em Quark gluon plasma} (R.~Hwa and X.~N. Wang,
  eds.).
\newblock World Scientific, 2003.
\newblock \href{http://arXiv.org/abs/hep-ph/0303204}{{\tt
  arXiv:hep-ph/0303204}}.

\bibitem{Weigert:2005us}
H.~Weigert,  \mbox{}  \href{http://dx.doi.org/10.1016/j.ppnp.2005.01.029}{{\em
  Prog. Part. Nucl. Phys.} {\bf 55} (2005) 461}
  [\href{http://arXiv.org/abs/hep-ph/0501087}{{\tt arXiv:hep-ph/0501087}}].

\bibitem{McLerran:1994ni}
L.~D. McLerran and R.~Venugopalan,  \mbox{}
  \href{http://dx.doi.org/10.1103/PhysRevD.49.2233}{{\em Phys. Rev.} {\bf D49}
  (1994) 2233} [\href{http://arXiv.org/abs/hep-ph/9309289}{{\tt
  arXiv:hep-ph/9309289}}].

\bibitem{McLerran:1994ka}
L.~D. McLerran and R.~Venugopalan,  \mbox{}
  \href{http://dx.doi.org/10.1103/PhysRevD.49.3352}{{\em Phys. Rev.} {\bf D49}
  (1994) 3352} [\href{http://arXiv.org/abs/hep-ph/9311205}{{\tt
  arXiv:hep-ph/9311205}}].

\bibitem{McLerran:1994vd}
L.~D. McLerran and R.~Venugopalan,  \mbox{}
  \href{http://dx.doi.org/10.1103/PhysRevD.50.2225}{{\em Phys. Rev.} {\bf D50}
  (1994) 2225} [\href{http://arXiv.org/abs/hep-ph/9402335}{{\tt
  arXiv:hep-ph/9402335}}].

\bibitem{Lappi:2006fp}
T.~Lappi and L.~McLerran,  \mbox{}
  \href{http://dx.doi.org/10.1016/j.nuclphysa.2006.04.001}{{\em Nucl. Phys.}
  {\bf A772} (2006) 200} [\href{http://arXiv.org/abs/hep-ph/0602189}{{\tt
  arXiv:hep-ph/0602189}}].

\bibitem{Kharzeev:2003wz}
D.~Kharzeev, Y.~V. Kovchegov and K.~Tuchin,  \mbox{}
  \href{http://dx.doi.org/10.1103/PhysRevD.68.094013}{{\em Phys. Rev.} {\bf
  D68} (2003) 094013} [\href{http://arXiv.org/abs/hep-ph/0307037}{{\tt
  arXiv:hep-ph/0307037}}].

\bibitem{Blaizot:2004wu}
J.~P. Blaizot, F.~Gelis and R.~Venugopalan,  \mbox{}
  \href{http://dx.doi.org/10.1016/j.nuclphysa.2004.07.005}{{\em Nucl. Phys.}
  {\bf A743} (2004) 13} [\href{http://arXiv.org/abs/hep-ph/0402256}{{\tt
  arXiv:hep-ph/0402256}}].

\bibitem{Gelis:2005pt}
F.~Gelis and Y.~Mehtar-Tani,  \mbox{}
  \href{http://dx.doi.org/10.1103/PhysRevD.73.034019}{{\em Phys. Rev.} {\bf
  D73} (2006) 034019} [\href{http://arXiv.org/abs/hep-ph/0512079}{{\tt
  arXiv:hep-ph/0512079}}].

\bibitem{Gelis:2006tb}
F.~Gelis, A.~M. Stasto and R.~Venugopalan,  \mbox{}
  \href{http://dx.doi.org/10.1140/epjc/s10052-006-0020-x}{{\em Eur. Phys. J.}
  {\bf C48} (2006) 489} [\href{http://arXiv.org/abs/hep-ph/0605087}{{\tt
  arXiv:hep-ph/0605087}}].

\bibitem{Kovchegov:2000hz}
Y.~V. Kovchegov,  \mbox{}
  \href{http://dx.doi.org/10.1016/S0375-9474(01)00652-2}{{\em Nucl. Phys.} {\bf
  A692} (2001) 557} [\href{http://arXiv.org/abs/hep-ph/0011252}{{\tt
  arXiv:hep-ph/0011252}}].

\bibitem{Blaizot:2008yb}
J.-P. Blaizot and Y.~Mehtar-Tani,  \mbox{}
  \href{http://dx.doi.org/doi:10.1016/j.nuclphysa.2008.11.010}{{\em Nucl.
  Phys.} {\bf A818} (2009) 97} [\href{http://arXiv.org/abs/0806.1422}{{\tt
  arXiv:0806.1422 [hep-ph]}}].

\bibitem{Krasnitz:1998ns}
A.~Krasnitz and R.~Venugopalan,  \mbox{}
  \href{http://dx.doi.org/10.1016/S0550-3213(99)00366-1}{{\em Nucl. Phys.} {\bf
  B557} (1999) 237} [\href{http://arXiv.org/abs/hep-ph/9809433}{{\tt
  arXiv:hep-ph/9809433}}].

\bibitem{Krasnitz:2000gz}
A.~Krasnitz and R.~Venugopalan,  \mbox{}
  \href{http://dx.doi.org/10.1103/PhysRevLett.86.1717}{{\em Phys. Rev. Lett.}
  {\bf 86} (2001) 1717} [\href{http://arXiv.org/abs/hep-ph/0007108}{{\tt
  arXiv:hep-ph/0007108}}].

\bibitem{Krasnitz:2001qu}
A.~Krasnitz, Y.~Nara and R.~Venugopalan,  \mbox{}
  \href{http://dx.doi.org/10.1103/PhysRevLett.87.192302}{{\em Phys. Rev. Lett.}
  {\bf 87} (2001) 192302} [\href{http://arXiv.org/abs/hep-ph/0108092}{{\tt
  arXiv:hep-ph/0108092}}].

\bibitem{Lappi:2003bi}
T.~Lappi,  \mbox{}  \href{http://dx.doi.org/10.1103/PhysRevC.67.054903}{{\em
  Phys. Rev.} {\bf C67} (2003) 054903}
  [\href{http://arXiv.org/abs/hep-ph/0303076}{{\tt arXiv:hep-ph/0303076}}].

\bibitem{Krasnitz:2003jw}
A.~Krasnitz, Y.~Nara and R.~Venugopalan,  \mbox{}
  \href{http://dx.doi.org/10.1016/j.nuclphysa.2003.08.004}{{\em Nucl. Phys.}
  {\bf A727} (2003) 427} [\href{http://arXiv.org/abs/hep-ph/0305112}{{\tt
  arXiv:hep-ph/0305112}}].

\bibitem{Lappi:2007ku}
T.~Lappi,  \mbox{}
  \href{http://dx.doi.org/10.1140/epjc/s10052-008-0588-4}{{\em Eur. Phys. J.}
  {\bf C55} (2008) 285} [\href{http://arXiv.org/abs/0711.3039}{{\tt
  arXiv:0711.3039 [hep-ph]}}].

\bibitem{Lappi:2009xa}
T.~Lappi, S.~Srednyak and R.~Venugopalan,  \mbox{}
  \href{http://dx.doi.org/10.1007/JHEP01(2010)066}{{\em JHEP} {\bf 01} (2010)
  066} [\href{http://arXiv.org/abs/0911.2068}{{\tt arXiv:0911.2068 [hep-ph]}}].

\bibitem{Kovner:1995ts}
A.~Kovner, L.~D. McLerran and H.~Weigert,  \mbox{}
  \href{http://dx.doi.org/10.1103/PhysRevD.52.3809}{{\em Phys. Rev.} {\bf D52}
  (1995) 3809} [\href{http://arXiv.org/abs/hep-ph/9505320}{{\tt
  arXiv:hep-ph/9505320}}].

\bibitem{Kovner:1995ja}
A.~Kovner, L.~D. McLerran and H.~Weigert,  \mbox{}
  \href{http://dx.doi.org/10.1103/PhysRevD.52.6231}{{\em Phys. Rev.} {\bf D52}
  (1995) 6231} [\href{http://arXiv.org/abs/hep-ph/9502289}{{\tt
  arXiv:hep-ph/9502289}}].

\bibitem{Dumitru:2001ux}
A.~Dumitru and L.~D. McLerran,  \mbox{}
  \href{http://dx.doi.org/10.1016/S0375-9474(01)01301-X}{{\em Nucl. Phys.} {\bf
  A700} (2002) 492} [\href{http://arXiv.org/abs/hep-ph/0105268}{{\tt
  arXiv:hep-ph/0105268}}].

\bibitem{Romatschke:2005pm}
P.~Romatschke and R.~Venugopalan,  \mbox{}
  \href{http://dx.doi.org/10.1103/PhysRevLett.96.062302}{{\em Phys. Rev. Lett.}
  {\bf 96} (2006) 062302} [\href{http://arXiv.org/abs/hep-ph/0510121}{{\tt
  arXiv:hep-ph/0510121}}].

\bibitem{Romatschke:2006nk}
P.~Romatschke and R.~Venugopalan,  \mbox{}
  \href{http://dx.doi.org/10.1103/PhysRevD.74.045011}{{\em Phys. Rev.} {\bf
  D74} (2006) 045011} [\href{http://arXiv.org/abs/hep-ph/0605045}{{\tt
  arXiv:hep-ph/0605045}}].

\bibitem{Krasnitz:2002ng}
A.~Krasnitz, Y.~Nara and R.~Venugopalan,  \mbox{}
  \href{http://dx.doi.org/10.1016/S0370-2693(02)03272-0}{{\em Phys. Lett.} {\bf
  B554} (2003) 21} [\href{http://arXiv.org/abs/hep-ph/0204361}{{\tt
  arXiv:hep-ph/0204361}}].

\bibitem{Krasnitz:2002mn}
A.~Krasnitz, Y.~Nara and R.~Venugopalan,  \mbox{}
  \href{http://dx.doi.org/10.1016/S0375-9474(03)00636-5}{{\em Nucl. Phys.} {\bf
  A717} (2003) 268} [\href{http://arXiv.org/abs/hep-ph/0209269}{{\tt
  arXiv:hep-ph/0209269}}].

\bibitem{Lappi:2006xc}
T.~Lappi and R.~Venugopalan,  \mbox{}
  \href{http://dx.doi.org/10.1103/PhysRevC.74.054905}{{\em Phys. Rev.} {\bf
  C74} (2006) 054905} [\href{http://arXiv.org/abs/nucl-th/0609021}{{\tt
  arXiv:nucl-th/0609021}}].

\bibitem{Armesto:2004ud}
N.~Armesto, C.~A. Salgado and U.~A. Wiedemann,  \mbox{}
  \href{http://dx.doi.org/10.1103/PhysRevLett.94.022002}{{\em Phys. Rev. Lett.}
  {\bf 94} (2005) 022002} [\href{http://arXiv.org/abs/hep-ph/0407018}{{\tt
  arXiv:hep-ph/0407018}}].

\bibitem{Hirano:2004rs}
T.~Hirano and Y.~Nara,  \mbox{}
  \href{http://dx.doi.org/10.1016/j.nuclphysa.2004.08.003}{{\em Nucl. Phys.}
  {\bf A743} (2004) 305} [\href{http://arXiv.org/abs/nucl-th/0404039}{{\tt
  arXiv:nucl-th/0404039}}].

\bibitem{Drescher:2006pi}
H.-J. Drescher, A.~Dumitru, A.~Hayashigaki and Y.~Nara,  \mbox{}
  \href{http://dx.doi.org/10.1103/PhysRevC.74.044905}{{\em Phys. Rev.} {\bf
  C74} (2006) 044905} [\href{http://arXiv.org/abs/nucl-th/0605012}{{\tt
  arXiv:nucl-th/0605012}}].

\bibitem{Drescher:2006ca}
H.~J. Drescher and Y.~Nara,  \mbox{}
  \href{http://dx.doi.org/10.1103/PhysRevC.75.034905}{{\em Phys. Rev.} {\bf
  C75} (2007) 034905} [\href{http://arXiv.org/abs/nucl-th/0611017}{{\tt
  arXiv:nucl-th/0611017}}].

\bibitem{Albacete:2007sm}
J.~L. Albacete,  \mbox{}
  \href{http://dx.doi.org/10.1103/PhysRevLett.99.262301}{{\em Phys. Rev. Lett.}
  {\bf 99} (2007) 262301} [\href{http://arXiv.org/abs/0707.2545}{{\tt
  arXiv:0707.2545 [hep-ph]}}].

\bibitem{Albacete:2003iq}
J.~L. Albacete, N.~Armesto, A.~Kovner, C.~A. Salgado and U.~A. Wiedemann,
  \mbox{}  \href{http://dx.doi.org/10.1103/PhysRevLett.92.082001}{{\em Phys.
  Rev. Lett.} {\bf 92} (2004) 082001}
  [\href{http://arXiv.org/abs/hep-ph/0307179}{{\tt arXiv:hep-ph/0307179}}].

\bibitem{Kharzeev:2000ph}
D.~Kharzeev and M.~Nardi,  \mbox{}
  \href{http://dx.doi.org/10.1016/S0370-2693(01)00457-9}{{\em Phys. Lett.} {\bf
  B507} (2001) 121} [\href{http://arXiv.org/abs/nucl-th/0012025}{{\tt
  arXiv:nucl-th/0012025}}].

\bibitem{Kharzeev:2001gp}
D.~Kharzeev and E.~Levin,  \mbox{}
  \href{http://dx.doi.org/10.1016/S0370-2693(01)01309-0}{{\em Phys. Lett.} {\bf
  B523} (2001) 79} [\href{http://arXiv.org/abs/nucl-th/0108006}{{\tt
  arXiv:nucl-th/0108006}}].

\bibitem{Gelis:2007kn}
F.~Gelis, T.~Lappi and R.~Venugopalan,  \mbox{}
  \href{http://dx.doi.org/10.1142/S0218301307008331}{{\em Int. J. Mod. Phys.}
  {\bf E16} (2007) 2595} [\href{http://arXiv.org/abs/0708.0047}{{\tt
  arXiv:0708.0047 [hep-ph]}}].

\bibitem{Gelis:2008ad}
F.~Gelis, T.~Lappi and R.~Venugopalan,  \mbox{}
  \href{http://dx.doi.org/10.1103/PhysRevD.78.054020}{{\em Phys. Rev.} {\bf
  D78} (2008) 054020} [\href{http://arXiv.org/abs/0807.1306}{{\tt
  arXiv:0807.1306 [hep-ph]}}].

\bibitem{Gelis:2008sz}
F.~Gelis, T.~Lappi and R.~Venugopalan,  \mbox{}
  \href{http://dx.doi.org/10.1103/PhysRevD.79.094017}{{\em Phys. Rev.} {\bf
  D79} (2008) 094017} [\href{http://arXiv.org/abs/0810.4829}{{\tt
  arXiv:0810.4829 [hep-ph]}}].

\bibitem{Dusling:2009ni}
K.~Dusling, F.~Gelis, T.~Lappi and R.~Venugopalan,  \mbox{}
  \href{http://dx.doi.org/10.1016/j.nuclphysa.2009.12.044}{{\em Nucl. Phys.}
  {\bf A836} (2010) 159} [\href{http://arXiv.org/abs/0911.2720}{{\tt
  arXiv:0911.2720 [hep-ph]}}].

\end{thebibliography}\endgroup
\bibliographystyle{JHEP-2mod}

\end{document}